\documentclass[prb,reprint,a4paper,superscriptaddress,showpacs,citeautoscript,floatfix]{revtex4-1}
\pdfoutput=1
\usepackage[charter]{mathdesign}
\usepackage{amsmath}
\usepackage[pdftex]{graphicx}
\usepackage{microtype}
\usepackage{bm}\let\vec\bm

\usepackage[svgnames]{xcolor}
\usepackage[
	colorlinks=True,linkcolor=DarkRed,citecolor=ForestGreen,urlcolor=MediumBlue,
	pdfstartview=FitH,bookmarks=False,pdfpagemode=UseNone
]{hyperref}

\begin{document}

\title{Second-order response theory of radio-frequency spectroscopy for cold atoms}

\author{C. Berthod}
\affiliation{Department of Quantum Matter Physics, University of Geneva, 24 quai Ernest-Ansermet, 1211 Geneva, Switzerland}
\author{M. K{\"o}hl}
\affiliation{Physikalisches Institut, University of Bonn, Wegelerstrasse 8, 53115 Bonn, Germany}
\author{T. Giamarchi}
\affiliation{Department of Quantum Matter Physics, University of Geneva, 24 quai Ernest-Ansermet, 1211 Geneva, Switzerland}

\date{June 1, 2015}

\begin{abstract}

We present a theoretical description of the radio-frequency (rf) spectroscopy of
fermionic atomic gases, based on the second-order response theory at finite
temperature. This approach takes into account the energy resolution due to the
envelope of the rf pulse. For a noninteracting final state, the momentum- and
energy-resolved rf intensity depends on the fermion spectral function and pulse
envelope. The contributions due to interactions in the final state can be
classified by means of diagrams. Using this formalism, as well as the local
density approximation in two and three dimensions, we study the interplay of
inhomogeneities and Hartree energy in forming the line shape of the rf signal.
We show that the effects of inhomogeneities can be minimized by taking advantage
of interactions in the final state, and we discuss the most relevant final-state
effects at low temperature and density, in particular the effect of a finite
lifetime.

\end{abstract}

\maketitle

\section{Introduction}

In many-fermion systems, the low-energy properties are often determined by
single-particle excitations across the Fermi surface. The character of these
excitations depends on the nature of the ground state, which itself depends on
the interactions. The study of single-particle excitations is therefore a key to
understanding the ground state and the role of interactions. In superconductors,
for instance, a gap in the single-particle excitation spectrum reveals the
condensation of Cooper pairs in the ground state. In a large class of materials,
the interactions bring only quantitative changes with respect to a
noninteracting ground state. The single-particle excitations are then similar to
uncorrelated particles, albeit with a renormalized mass and a finite lifetime.
The collection of these ``quasiparticles'' forms a Fermi liquid, which can be
characterized by a small number of effective parameters \cite{*[] [{ [Sov. Phys.
JETP \textbf{3}, 920 (1957)].}] Landau-1956}. For strong interactions and/or
reduced dimensionality, qualitative changes may occur in the ground state,
leading to the disappearance of the quasiparticles and the emergence of more
complex, sometimes mysterious, excitations \cite{Stewart-2001, Schofield-2009}.
The absence of quasiparticles in a fermion system is a hallmark of non-Fermi
liquid physics, indicating an unconventional ground state.

For electronic materials, angle-resolved photoemission spectroscopy (ARPES)
gives access to the single-particle excitations and allows one to probe the
existence of quasiparticles \cite{Damascelli-2003}. The signature of
quasiparticles is a peak at low energy in the spectral function, which is the
momentum-energy distribution of the single-particle excitations, denoted
$A(\vec{k},\varepsilon)$. Conversely, a structureless spectral function signals
the absence of quasiparticles. ARPES experiments require clean surfaces and
ultrahigh vacuum, and an energy resolution below the typical excitation energy
of the quasiparticles. Steady improvements in recent years and the development
of laser ARPES have made it possible to measure the spectral function with
excellent resolution in several condensed-matter systems \cite{Vishik-2010,
Lu-2012, Allan-2013}. When it is present, the quasiparticle peak and its
dispersion anomalies can help in identifying the interactions that determine the
quasiparticle dynamics.

Fermionic cold-atom gases open new avenues in the study of quasiparticles,
especially thanks to the possibility of tuning both the dimensionality and the
strength of interactions. Radio-frequency (rf) spectroscopy is presently the
best method to measure the spectral function of cold-atom systems. Unlike in
conventional ARPES, photoemission spectroscopy in ultracold atoms is performed
using rf photons, which carry negligible momentum but only supply an energy
$h\nu$. The momentum of the extracted atoms is then measured using the
time-of-flight technique. If the particles are decoupled in the final state,
their energy and momentum distributions follow the spectral function of the
photon-induced hole, which is the occupied part of the spectral function, i.e.,
$A(\vec{k},\varepsilon)f(\varepsilon)$, where $f(\varepsilon)$ is the Fermi
function \cite{Torma-2000, Dao-2007, *Dao-2009}.

The interpretation of photoemission and rf experiments may be complicated by the
unavoidable interaction in the final state, as well as several other
difficulties. In ARPES, these are, for instance, the sample surface, which
breaks inversion symmetry and produces interference, or the screening of the
electromagnetic field, which prevents light from entering the bulk of the
material. In rf spectroscopy of cold atoms, the main concern is the
inhomogeneity of harmonically trapped gases. When interactions and excitation
energies are not too low, as in the studies of the BCS-BEC crossover
\cite{Ketterle-2008}, some of these difficulties may turn out to be irrelevant.
For weak interactions, however, they will eventually become important. If the
signal is broadened by final-state effects, averaging over inhomogeneities, and
finite energy resolution, a precise modeling is necessary in order to recover
the crucial information about the quasiparticles.

In the established theory of rf spectroscopy, one computes the instantaneous
transition rate to the final state. This can be done either by linear response
\cite{Torma-2000}, which provides the current $\dot{N}_f$ of particles
transferred to the final state, or by time-dependent perturbation theory (Fermi
golden rule) \cite{Dao-2007, *Dao-2009}. At leading order, $\dot{N}_f$ is
related to a response function, which can be represented by bubblelike Feynman
diagrams \cite{Perali-2008}. In this approach, the effect of inhomogeneities has
been investigated at the mean-field level \cite{Ohashi-2005} or using the
local-density approximation (LDA) \cite{He-2005, Dao-2007, *Dao-2009}. To
circumvent the difficulties raised by inhomogeneity, a Raman local spectroscopy
was proposed theoretically \cite{Dao-2007, *Dao-2009}, while a tomographic
technique \cite{Shin-2007} and a method to selectively address the cloud center
\cite{Drake-2012, *Sagi-2012} were demonstrated. Final-state effects have been
treated in the mean-field approximation \cite{Yu-2006} by sum-rule arguments
\cite{Baym-2007, Punk-2007}, within a reduced basis \cite{Basu-2008}, a $1/N$
expansion \cite{Veillette-2008}, diagrammatically \cite{Perali-2008,
Pieri-2009}, or through self-consistency requirements \cite{He-2009}. Most of
these studies have focussed on the BCS-BEC crossover problem.

For intermediate or weak interactions, the finite energy resolution must be
considered. The relevant quantity to calculate is no longer $\dot{N}_f$, but the
total population $N_f$ of the final state, created over the duration of the rf
pulse. Momentum-resolved rf experiments indeed measure the momentum distribution
$n_f(\vec{k},t)$ at a time $t$ after the extinction of the rf pulse. If atoms
were excited at a constant rate, $\dot{N}_f$ and $N_f$ would carry the same
information, but this is not the case in practice. In this paper, we present the
calculation of $n_f(\vec{k},t)$ within equilibrium response theory. The
derivation is performed in the finite-temperature Matsubara framework. Unlike
$\dot{N}_f$, $n_f$ vanishes at first order in the atom-light coupling. At second
order, the momentum distribution is related to a three-point response function,
whose contributions can be classified using Feynman diagrams. These diagrams
have three external vertices, unlike the bubble diagrams of the established
theory, which have only two. The leading contribution reproduces the known
result \cite{Torma-2000, Dao-2007, *Dao-2009}, albeit convolved with a
resolution function, which depends on the envelope of the rf pulse and on the
spectral function in the final state.

Simulations based on this formalism have been presented earlier
\cite{Frohlich-2012} and compared with measurements for $^{40}$K atoms in
two-dimensional harmonic traps with a weak attractive interaction. In this work,
it was shown that inhomogeneities must be considered for a correct determination
of the quasiparticle effective mass. Here we discuss the role of inhomogeneities
in this experiment in more detail and propose ways to reduce their effect. We
also show that, in the experiments of Ref.~\onlinecite{Gupta-2003} made with
$^6$Li atoms in three-dimensional harmonic traps, the inhomogeneity sets the
line shape of the integrated rf intensity and should be considered for the
precise experimental determination of the scattering length.

The paper is organized as follows. In Sec.~\ref{sec:model}, we present the model
and the calculation of $n_f(\vec{k},t)$. The generic diagrams giving the
momentum distribution are shown in Sec.~\ref{sec:diagrams}, and the leading
contribution is evaluated in Sec.~\ref{sec:leading_contribution}. Without
interaction in the final state, the analysis simplifies as shown in
Sec.~\ref{sec:no-interaction}, where we discuss the interplay between the
inhomogeneity and the Hartree shifts. We study final-state effects in
Sec.~\ref{sec:final-state}: the effect of a finite lifetime, the effect of
inhomogeneous Hartree shifts, and other final-state effects which correspond
diagrammatically to vertex corrections. Conclusions and perspectives are given
in Sec.~\ref{sec:conclusion}.

\section{Finite-temperature, second-order response theory for the momentum distribution}
\label{sec:theory}

\subsection{Description of the model}
\label{sec:model}

\begin{figure}[b]
\includegraphics[width=0.4\columnwidth]{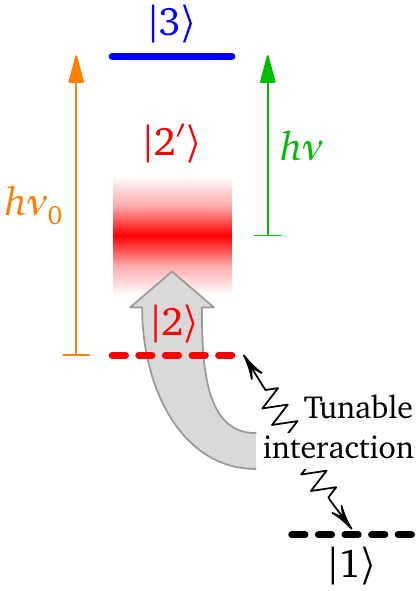}
\caption{\label{fig:fig01}(Color online)
Rf spectroscopy of ultracold atoms. The hyperfine atomic levels $|1\rangle$,
$|2\rangle$, and $|3\rangle$ are split by the Zeeman effect, such that a rf
transition is allowed between states $|2\rangle$ and $|3\rangle$. The tunable
many-body interaction between atoms of the cloud in states $|1\rangle$ and
$|2\rangle$ shifts and broadens the level $|2\rangle$. The effects of this
interaction are probed by comparing the frequency $\nu$ of the rf transition
with the frequency $\nu_0$ of the noninteracting transition. Spin relaxation
from $|2\rangle$ to $|1\rangle$ is forbidden by conservation of spin, meaning
that the energy splitting between $|1\rangle$ and $|2\rangle$ is irrelevant.
Levels $|1\rangle$ and $|3\rangle$ may be renormalized by interactions as well.
}
\end{figure}

The atoms are modeled as three-level systems with internal states
$|\alpha\rangle$, $\alpha=1,2,3$. States $|1\rangle$ and $|2\rangle$ are assumed
to interact most strongly, while state $|3\rangle$ has higher energy and will be
the final state of the rf experiment; see Fig.~\ref{fig:fig01}. We consider that
the three levels have the same dispersion; our results are readily generalized
to the case where the dispersions are different in the initial and final states.
These levels correspond, in practice, to atomic hyperfine states. The
interaction between $|1\rangle$ and $|2\rangle$ is resonant and can be tuned by
means of a Feshbach resonance \cite{Chin-2010}. Once the field is set, the
interactions between $|1\rangle$ and $|3\rangle$ and between $|2\rangle$ and
$|3\rangle$ are also set. Ideally, the latter interactions are small compared
with the former. We consider hereafter fermionic atoms, and we assume
translation invariance for simplicity. The extension to bosons is
straightforward, and the formalism can be developed in real space if needed. Let
$c^{\dagger}_{\alpha\vec{k}}$ be the creation operator for an atom in the state
$|\alpha\rangle$ with momentum $\hbar\vec{k}$. The low-energy effective
Hamiltonian is $H=H_0+V$, with
	\begin{align}\label{eq:H0}
		H_0&=\sum_{\vec{k}}\left[\varepsilon_{\vec{k}}(c^{\dagger}_{1\vec{k}}
		c^{\phantom{\dagger}}_{1\vec{k}}+c^{\dagger}_{2\vec{k}}
		c^{\phantom{\dagger}}_{2\vec{k}})+(\varepsilon_{\vec{k}}+h\nu_0)
		c^{\dagger}_{3\vec{k}}c^{\phantom{\dagger}}_{3\vec{k}}\right]\\
		\label{eq:V}
		V&=\frac{1}{2}\sum_{\stackrel{\scriptstyle\vec{k}\vec{k}'\vec{q}}
		{\alpha\beta}}V_{\alpha\beta}(\vec{q})c^{\dagger}_{\alpha\vec{k}}
		c^{\dagger}_{\beta\vec{q}-\vec{k}}c^{\phantom{\dagger}}_{\alpha\vec{k}'}
		c^{\phantom{\dagger}}_{\beta\vec{q}-\vec{k}'}.
	\end{align}
We consider here the case of a local interaction between two atoms with
center-of-mass momentum $\vec{q}$. The detailed form of the interaction plays no
role in our derivation, which is also valid for more general momentum-dependent
interactions. For a contact interaction, the Pauli principle prevents atoms in
the same internal state from interacting, and we can set $V_{\alpha\alpha}=0$.

The level separation $\nu_0$ is typically in the 100 MHz range, and the rf
radiation at this frequency has a wavelength of the order of meters. The rf
pulse therefore induces momentum-conserving transitions. Let $H'(t)$ be the
time-dependent interaction between the rf radiation and the atoms, and let us
assume that the allowed transition is between states $|2\rangle$ and
$|3\rangle$. We have
	\begin{equation}\label{eq:conversion}
		H'(t)=\mathcal{E}(t)\sum_{\vec{k}}\left(c^{\dagger}_{3\vec{k}}
		c^{\phantom{\dagger}}_{2\vec{k}}+\text{h.c.}\right)=
		\mathcal{E}(t)\sum_{\vec{k}}\gamma^{\phantom{\dagger}}_{\vec{k}}.
	\end{equation}
The function $\mathcal{E}(t)$ gives the time envelope and strength of the
coupling. For later convenience, we define the operator
	\begin{equation}
		\gamma^{\phantom{\dagger}}_{\vec{k}}=\gamma^{\dagger}_{\vec{k}}
		=c^{\dagger}_{3\vec{k}}
		c^{\phantom{\dagger}}_{2\vec{k}}+\text{h.c.}
	\end{equation}

\subsection{Generic diagrams for the momentum distribution}
\label{sec:diagrams}

We now expand the momentum distribution in the final state, $n_{\vec{k}}\equiv
c^{\dagger}_{3\vec{k}}c^{\phantom{\dagger}}_{3\vec{k}}$, in powers of $H'$. In
the grand-canonical ensemble, and in the interaction picture, we have
	\begin{equation}
		\langle n_{\vec{k}}(t)\rangle=\text{Tr}\,\rho n_{\vec{k}}(t),
	\end{equation}
with $\rho=e^{-\beta(H-\mu N)}/\text{Tr}\,e^{-\beta(H-\mu N)}$,
$\beta=1/(k_{\text{B}}T)$, $\mu$ the chemical potential, and $N$ the number
operator. Since we work in equilibrium, the chemical potential $\mu$ sets the
populations of the three levels, and $N$ is the total atom number. The evolution
is given by $n_{\vec{k}}(t)=U^{-1}(t)n_{\vec{k}}U(t)$ with $U(t)$ the
interaction part of the evolution operator. The zeroth-order term is obviously
	\begin{equation}\label{eq:nk0}
		\langle n_{\vec{k}}(t)\rangle^{(0)}=\text{Tr}\,\rho n_{\vec{k}}
		\equiv\langle n_{\vec{k}}\rangle_H,
	\end{equation}
which gives the equilibrium thermal population of the final state. For a
noninteracting system, this becomes
	\begin{equation}\label{eq:nk00}
		\langle n_{\vec{k}}\rangle_{H_0}=f(\varepsilon_{\vec{k}}+h\nu_0-\mu),
	\end{equation}
where $f(\varepsilon)=(e^{\beta\varepsilon}+1)^{-1}$ is the Fermi distribution
function. The usual setup is that the final state is initially empty, such that
this contribution is negligible at low temperatures, $k_{\text{B}}T\ll h\nu_0$.
The first-order term in $H'$ is known, from standard linear-response theory, to
be
	\begin{align*}
		\langle n_{\vec{k}}(t)\rangle^{(1)}&=-\frac{i}{\hbar}\int_{-\infty}^tdt'\,
		\langle[n_{\vec{k}}(t),H'(t')]\rangle_H\\
		&=\int_{-\infty}^{\infty}
		\frac{d\omega}{2\pi}\,e^{-i\omega t}\mathscr{E}(\omega)\sum_{\vec{k'}}
		C^R_{n_{\vec{k}}\gamma_{\vec{k}'}}(\omega)=0.
	\end{align*}
At the second line, we have introduced $\mathscr{E}(\omega)$, the Fourier
transform of $\mathscr{E}(t)$, and the equilibrium retarded correlation function
of the operators $n_{\vec{k}}$ and $\gamma_{\vec{k}'}$ in the system described
by $H$. In the time domain, this correlation function is
	\begin{equation}\label{eq:Cngamma}
		C^R_{n_{\vec{k}}\gamma_{\vec{k}'}}(t)=-\frac{i}{\hbar}\theta(t)
		\langle[n_{\vec{k}}(t),\gamma_{\vec{k}'}(0)]\rangle_H,
	\end{equation}
with the time dependence of the operators governed by the evolution
$e^{-i(H-\mu N)t/\hbar}$. Because $H$, $N$, and $n_{\vec{k}}$ all conserve the
number of atoms in the states $|2\rangle$ and $|3\rangle$, while the two terms
in $\gamma_{\vec{k}'}$ do not, the correlation function (\ref{eq:Cngamma})
vanishes identically.

\begin{figure}[tb]
\includegraphics[width=\columnwidth]{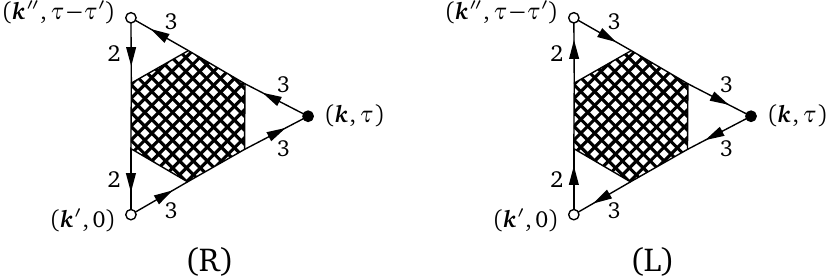}
\caption{\label{fig:fig02}
Diagrammatic representation of the double-imaginary time function
(\ref{eq:C2definition}). Diagrams (R) and (L) correspond to the first and second
terms of Eq.~(\ref{eq:C2terms}), respectively, up to a minus sign (the
correlation function equals minus the diagram, if standard diagrammatic rules
are used). The two diagrams are identical, except for the direction of the
arrows (R, right-handed, L, left-handed). The white dots denote the interaction
of atoms with light, leading to transitions between internal states $|2\rangle$
and $|3\rangle$; the black dots denote the measured momentum distribution, and
the hatched regions represent all interactions, including interactions with the
state $|1\rangle$.
}
\end{figure}

The second-order response involves a double commutator and can be expressed in
terms of a double-time retarded correlation function of the three operators
$n_{\vec{k}}$, $\gamma_{\vec{k}'}$, and $\gamma_{\vec{k}''}$:
	\begin{align}
		\nonumber
		\langle n_{\vec{k}}(t)\rangle^{(2)}&=
		\left(-\frac{i}{\hbar}\right)^2\int_{-\infty}^tdt'
		\int_{-\infty}^{t'}dt''\\
		\nonumber
		&\hspace{1.5cm}\times\langle[[n_{\vec{k}}(t),H'(t')],H'(t'')]\rangle_H\\
		\nonumber
		&=\int_{-\infty}^{\infty}
		\frac{d\omega}{2\pi}\frac{d\omega'}{2\pi}\,e^{-i(\omega+\omega')t}
		\mathscr{E}(\omega)\mathscr{E}(\omega')\\
		\label{eq:nk2}
		&\hspace{1.5cm}\times\sum_{\vec{k}'\vec{k}''}
		C^R_{n_{\vec{k}}\gamma_{\vec{k}'}\gamma_{\vec{k}''}}(\omega,\omega').
	\end{align}
This contribution can be evaluated within the Matsubara formalism. We find that
the double-time correlation function in Eq.~(\ref{eq:nk2}) is given by the
analytic continuation to real frequencies of an imaginary-frequency function
$\mathscr{C}_{n_{\vec{k}}\gamma_{\vec{k}'}\gamma_{\vec{k}''}}
(i\Omega_n,i\Omega_n')$, according to (see
Appendix~\ref{app:Matsubara-double-time})
	\begin{multline}\label{eq:C2continuation}
		C^R_{n_{\vec{k}}\gamma_{\vec{k}'}\gamma_{\vec{k}''}}(\omega,\omega')\\
		=\frac{1}{2}\mathscr{C}_{n_{\vec{k}}\gamma_{\vec{k}'}\gamma_{\vec{k}''}}
		(i\Omega_n\to\hbar\omega+i0^+,i\Omega_n'\to\hbar\omega'+i0^+).
	\end{multline}
$i\Omega_n=2n\pi k_{\text{B}}T$ with integer $n$ denote the even Matsubara
frequencies. In the imaginary-time domain, the double-time function is defined as
	\begin{equation}\label{eq:C2definition}
		\mathscr{C}_{n_{\vec{k}}\gamma_{\vec{k}'}\gamma_{\vec{k}''}}(\tau,\tau')=
		\langle T_{\tau}n_{\vec{k}}(\tau)\gamma_{\vec{k}'}(0)
		\gamma_{\vec{k}''}(\tau-\tau')\rangle_H,
	\end{equation}
with $T_{\tau}$ the imaginary-time ordering operator,
$n_{\vec{k}}(\tau)=e^{\tau(H-\mu N)}n_{\vec{k}}e^{-\tau(H-\mu N)}$, and
similarly for $\gamma_{\vec{k}''}(\tau-\tau')$. The imaginary-time and
imaginary-frequency functions are related by
	\begin{equation}\label{eq:C2Fourier}
		\mathscr{C}_{n_{\vec{k}}\gamma_{\vec{k}'}\gamma_{\vec{k}''}}(i\Omega_n,i\Omega_n')
		=\int_0^{\beta}d\tau d\tau'\,e^{i\Omega_n\tau}e^{i\Omega_n'\tau'}
		\mathscr{C}_{n_{\vec{k}}\gamma_{\vec{k}'}\gamma_{\vec{k}''}}(\tau,\tau').
	\end{equation}
The correlation function (\ref{eq:C2definition}) is nonzero, because the two
crossed terms in the product $\gamma_{\vec{k}'}\gamma_{\vec{k}''}$ conserve the
number of atoms of each flavor. Gathering these two terms, we get
	\begin{multline}\label{eq:C2terms}
		\mathscr{C}_{n_{\vec{k}}\gamma_{\vec{k}'}\gamma_{\vec{k}''}}(\tau,\tau')=\\
		\langle T_{\tau}c^{\dagger}_{3\vec{k}}(\tau)c^{\phantom{\dagger}}_{3\vec{k}}(\tau)
		c^{\dagger}_{3\vec{k}'}(0)c^{\phantom{\dagger}}_{2\vec{k}'}(0)
		c^{\dagger}_{2\vec{k}''}(\tau-\tau')
		c^{\phantom{\dagger}}_{3\vec{k}''}(\tau-\tau')\rangle_H\\
		+\langle T_{\tau}c^{\dagger}_{3\vec{k}}(\tau)c^{\phantom{\dagger}}_{3\vec{k}}(\tau)
		c^{\dagger}_{2\vec{k}'}(0)c^{\phantom{\dagger}}_{3\vec{k}'}(0)
		c^{\dagger}_{3\vec{k}''}(\tau-\tau')
		c^{\phantom{\dagger}}_{2\vec{k}''}(\tau-\tau')\rangle_H.\\[-1em]
	\end{multline}\\[-1.5em]
The two terms can be represented by Feynman diagrams, as shown in
Fig.~\ref{fig:fig02}. These diagrams have three entry points, one representing
the measurement of the momentum distribution and two representing the
transitions induced by $H'$ between states $|2\rangle$ and $|3\rangle$. Similar
diagrams arise in the response theory of electron photoemission
\cite{Schaich-1970, Caroli-1973, Chang-1973, Keiter-1978, Almbladh-2006}. This
is to be contrasted with the bubble-type diagrams representing the transition
rate $\dot{N}_f$ (see, e.g., Ref.~\onlinecite{Perali-2008}).

\subsection{Leading contribution}
\label{sec:leading_contribution}

\begin{figure}[t]
\includegraphics[width=0.9\columnwidth]{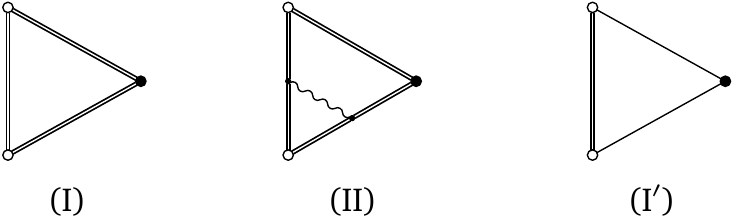}
\caption{\label{fig:fig03}
Leading term (I) and an example of vertex correction (II). In both cases, the
two topologically inequivalent diagrams of kinds R and L must be considered. For
a noninteracting final state, diagrams of type I$'$ are the only nonvanishing
contributions. Double lines denote the Green's function, and single lines denote
the noninteracting Green's function in the final state.
}
\end{figure}

We can distinguish two categories of diagrams, as illustrated in
Fig.~\ref{fig:fig03}. The justification for separating the diagrams of type I
from ``vertex corrections'' of type II stems from the fact that, in usual
experimental conditions, the interactions in the final state are small compared
with the other interactions. If the former are exactly zero
($V_{13}=V_{23}=V_{33}=0$), all vertex corrections of type II disappear. Then,
the two diagrams of type I$'$ (right- and left-handed) are the only nonvanishing
terms, with the two propagators in state $3$ given by free propagators. Since
diagram (I$'$) can be derived from diagram (I) by taking the appropriate limit,
we shall evaluate here diagram (I) and discuss the case of a noninteracting
final state in the next section. Final-state effects are present in both type-I
and type-II diagrams. We discuss final-state effects of type I in
Secs.~\ref{sec:final-state1} and \ref{sec:final-state2} and those of type II in
Sec.~\ref{sec:final-state3}.

The two diagrams of type I involve a single momentum, i.e.,
$\mathscr{C}_{n_{\vec{k}}\gamma_{\vec{k}'}\gamma_{\vec{k}''}}^{\text{(I)}}(\tau,
\tau')\propto\delta_{\vec{k}\vec{k}'}\delta_{\vec{k}\vec{k}''}$. They can be
expressed in terms of the Green's functions for each atomic state. We define the
fermionic Green's function as $\mathscr{G}_{\alpha}(\vec{k},\tau)= -\langle
T_{\tau}c^{\phantom{\dagger}}_{\alpha\vec{k}}(\tau)
c^{\dagger}_{\alpha\vec{k}}(0)\rangle_H$. In imaginary frequency, they are
	\begin{align}
		\mathscr{G}_{1,2}(\vec{k},i\omega_n)&=\frac{1}{i\omega_n-\xi_{\vec{k}}
		-\Sigma_{1,2}(\vec{k},i\omega_n)}\\
		\mathscr{G}_3(\vec{k},i\omega_n)&=\frac{1}{i\omega_n-\xi_{\vec{k}}-h\nu_0
		-\Sigma_3(\vec{k},i\omega_n)},
	\end{align}
where $i\omega_n=(2n+1)\pi k_{\text{B}}T$,
$\xi_{\vec{k}}=\varepsilon_{\vec{k}}-\mu$, and
$\Sigma_{\alpha}(\vec{k},i\omega_n)$ is the self-energy. Translating the two
diagrams using the conventions of Fig.~\ref{fig:fig02} gives
	\begin{multline}
		\mathscr{C}_{n_{\vec{k}}\gamma_{\vec{k}}\gamma_{\vec{k}}}^{\text{(I)}}(\tau,\tau')=
		\mathscr{G}_2(\vec{k},\tau'-\tau)
		\mathscr{G}_3(\vec{k},\tau)\mathscr{G}_3(\vec{k},-\tau')\\
		+\mathscr{G}_2(\vec{k},\tau-\tau')
		\mathscr{G}_3(\vec{k},\tau')\mathscr{G}_3(\vec{k},-\tau).
	\end{multline}
The minus sign associated with the fermion loop is canceled because the
correlation function equals minus the diagram. We perform the Fourier transform
in Eq.~(\ref{eq:C2Fourier}) using the spectral representation of the Green's
function,
	\begin{equation}
		\mathscr{G}_{\alpha}(\vec{k},\tau)=\int_{-\infty}^{\infty} d\varepsilon\,
		A_{\alpha}(\vec{k},\varepsilon)
		\frac{1}{\beta}\sum_{i\omega_n}\frac{e^{-i\omega_n\tau}}
		{i\omega_n-\varepsilon},
	\end{equation}
where
	$
		A_{\alpha}(\vec{k},\varepsilon)=-\text{Im}\,
		\mathscr{G}_{\alpha}(\vec{k},i\omega_n\to\varepsilon+i0^+)/\pi
	$
is the single-particle spectral function. This leads to
	\begin{multline}\label{eq:CnggI}
		\mathscr{C}_{n_{\vec{k}}\gamma_{\vec{k}}\gamma_{\vec{k}}}^{\text{(I)}}
		(i\Omega_n,i\Omega_n')=\int_{-\infty}^{\infty} d\varepsilon d\varepsilon'd\varepsilon''\\
		\times A_2(\vec{k},\varepsilon)\,A_3(\vec{k},\varepsilon')\,A_3(\vec{k},\varepsilon'')\\
		\times\frac{1}{\beta}\sum_{i\omega_n}\frac{1}{i\omega_n-\varepsilon}\left[
		\frac{1}{i\omega_n+i\Omega_n-\varepsilon'}\frac{1}{i\omega_n-i\Omega_n'-\varepsilon''}
		\right.\\\left.
		+\frac{1}{i\omega_n-i\Omega_n-\varepsilon'}\frac{1}{i\omega_n+i\Omega_n'-\varepsilon''}
		\right].
	\end{multline}
The frequency sums are evaluated in the usual manner \cite{Mahan-2000} and yield
terms proportional to either $f(\varepsilon)$, $f(\varepsilon')$, or
$f(\varepsilon'')$. For weak interactions in the final state,
$A_3(\vec{k},\varepsilon)$ peaks near $\varepsilon=\xi_{\vec{k}}+h\nu_0$. The
terms proportional to $f(\varepsilon')$, and $f(\varepsilon'')$ are therefore
small at low temperature, like the zeroth-order term (\ref{eq:nk0}). We denote
$\mathscr{C}_{n_{\vec{k}}\gamma_{\vec{k}}\gamma_{\vec{k}}}^{\text{(Ia)}}$ the
contribution of the dominant terms proportional to $f(\varepsilon)$ and
$\mathscr{C}_{n_{\vec{k}}\gamma_{\vec{k}}\gamma_{\vec{k}}}^{\text{(Ib)}}$ the
contribution of the other terms. We have
	\begin{multline}\label{eq:CnggIa}
		\hspace{-3mm}\mathscr{C}_{n_{\vec{k}}\gamma_{\vec{k}}\gamma_{\vec{k}}}^{\text{(Ia)}}
		(i\Omega_n,i\Omega_n')=\int_{-\infty}^{\infty} d\varepsilon d\varepsilon'd\varepsilon''\\
		\times A_2(\vec{k},\varepsilon)\,A_3(\vec{k},\varepsilon')\,A_3(\vec{k},\varepsilon'')\\
		\times f(\varepsilon)\left[
		\frac{1}{i\Omega_n+\varepsilon-\varepsilon'}\frac{-1}{i\Omega_n'-\varepsilon+\varepsilon''}
		\right.\\\left.
		+\frac{-1}{i\Omega_n-\varepsilon+\varepsilon'}\frac{1}{i\Omega_n'+\varepsilon-\varepsilon''}
		\right].
	\end{multline}
Making the analytic continuation as in Eq.~(\ref{eq:C2continuation}), and
inserting in Eq.~(\ref{eq:nk2}), we obtain the leading contribution to the
momentum distribution:
	\begin{multline}\label{eq:nkIa-0}
		\langle n_{\vec{k}}(t)\rangle^{\text{(Ia)}}=\int_{-\infty}^{\infty} d\varepsilon\,
		A_2(\vec{k},\varepsilon)\,f(\varepsilon)\\
		\times\scalebox{1}[1.3]{\Big|}\int_{-\infty}^{\infty}\!\!
		d\varepsilon'\,A_3(\vec{k},\varepsilon')
		\mathscr{F}_t(\varepsilon-\varepsilon')\scalebox{1}[1.3]{\Big|}^2.
	\end{multline}
The dimensionless function $\mathscr{F}$ accounts for the broadening effect due
to the rf pulse:
	\begin{equation}
		\mathscr{F}_t(\varepsilon)=\int_{-\infty}^{\infty}\frac{d\omega}{2\pi}
		\frac{e^{-i\omega t}\mathscr{E}(\omega)}{\hbar\omega-\varepsilon+i0^+}
		=\frac{-i}{\hbar}\int_{-\infty}^tdt'\,e^{i\frac{\varepsilon}{\hbar}(t'-t)}\mathscr{E}(t').
	\end{equation}
The main goal of rf spectroscopy is to determine the spectral function
$A_2(\vec{k},\varepsilon)$. For weak interactions, this function is peaked near
$\varepsilon=\varepsilon_{\vec{k}}-\mu=\xi_{\vec{k}}$. On the other hand, since
the dispersions in the initial and final states are the same and only
$\vec{q}=0$ transitions are possible, one expects to observe, by varying the
frequency $\nu$ of the rf radiation, a signal peaking close to the frequency
$\nu_0$ of the noninteracting transition. In order to make this more apparent,
we introduce the detuning $\tilde{\nu}=\nu_0-\nu$, we change variables in
Eq.~(\ref{eq:nkIa-0}), and rewrite it in the form
	\begin{subequations}\label{eq:nkIa}
	\begin{multline}
		\langle n_{\vec{k}}(t)\rangle^{\text{(Ia)}}=\int_{-\infty}^{\infty} d\varepsilon\,
		\mathscr{R}_{\vec{k}}^{\text{(Ia)}}(\varepsilon)\,
		A_2(\vec{k},\xi_{\vec{k}}+h\tilde{\nu}-\varepsilon)\\
		\times f(\xi_{\vec{k}}+h\tilde{\nu}-\varepsilon).
	\end{multline}
This shows that the measured momentum distribution is the convolution of the
occupied part of the spectral function with a dimensionless resolution function
$\mathscr{R}_{\vec{k}}^{\text{(Ia)}}$. Under ideal conditions, the resolution
function is proportional to $\delta(\varepsilon)$, and the momentum distribution
peaks near $\nu=\nu_0$, as expected. The expression of the resolution function
resulting from Eq.~(\ref{eq:nkIa-0}) is
	\begin{equation}\label{eq:RIa}
		\mathscr{R}_{\vec{k}}^{\text{(Ia)}}(\varepsilon)=\scalebox{1}[1.3]{\Big|}
		\int_{-\infty}^{\infty}
		d\varepsilon'\,A_3(\vec{k},\varepsilon')\mathscr{F}_t(\varepsilon+\varepsilon'
		-\xi_{\vec{k}}-h\tilde{\nu})\scalebox{1}[1.3]{\Big|}^2.
	\end{equation}
	\end{subequations}
It takes into account the renormalization of the final state by interactions, as
well as the broadening due to the time envelope of the rf pulse. For a
noninteracting final state with a spectral function
$A_3(\vec{k},\varepsilon)=\delta(\varepsilon-\xi_{\vec{k}}-h\nu_0)$, the
resolution function simplifies to
	\begin{equation}\label{eq:RIp}
		\mathscr{R}^{\text{(I$'$)}}(\varepsilon)=\frac{1}{\hbar^2}
		\scalebox{1}[1.3]{\Big|}
		\int_{-\infty}^tdt'\,e^{i(\varepsilon+h\nu)t'/\hbar}\mathscr{E}(t')
		\scalebox{1}[1.3]{\Big|}^2.
	\end{equation}
Equations~(\ref{eq:nkIa}) are one central result of this work. We use them to
study the interplay of Hartree shifts and inhomogeneity in two-dimensional
$^{40}$K (Sec.~\ref{sec:mH}) and three-dimensional $^6$Li (Sec.~\ref{sec:Li6})
and to study the effect of a finite lifetime in the final state
(Sec.~\ref{sec:final-state1}).

The terms resulting from the frequency sum in Eq.~(\ref{eq:CnggI}), which have
not been retained in Eq.~(\ref{eq:CnggIa}), are
	\begin{multline}
		\hspace{-3mm}\mathscr{C}_{n_{\vec{k}}\gamma_{\vec{k}}\gamma_{\vec{k}}}^{\text{(Ib)}}
		(i\Omega_n,i\Omega_n')=\int_{-\infty}^{\infty} d\varepsilon d\varepsilon'd\varepsilon''\\
		\times A_2(\vec{k},\varepsilon)A_3(\vec{k},\varepsilon')A_3(\vec{k},\varepsilon'')\\
		f(\varepsilon')\left[
		\left(
		\frac{1}{i\Omega_n+\varepsilon-\varepsilon'}+\frac{1}{i\Omega_n'+\varepsilon-\varepsilon'}
		\right)\frac{1}{i\Omega_n+i\Omega_n'-\varepsilon'+\varepsilon''}
		\right.\\ \left.
		+\left(
		\frac{1}{i\Omega_n-\varepsilon+\varepsilon'}+
		\frac{1}{i\Omega_n'-\varepsilon+\varepsilon'}\right)
		\frac{1}{i\Omega_n+i\Omega_n'+\varepsilon'-\varepsilon''}
		\right].
	\end{multline}
We have rearranged the terms by exchanging $\varepsilon'$ and $\varepsilon''$ in
half of them. We proceed as above, and introduce again a resolution function:
	\begin{subequations}\label{eq:nkIb}
	\begin{multline}
		\langle n_{\vec{k}}(t)\rangle^{\text{(Ib)}}=-f(\xi_{\vec{k}}+h\nu_0)
		\int_{-\infty}^{\infty} d\varepsilon\,\mathscr{R}_{\vec{k}}^{\text{(Ib)}}(\varepsilon)\\
		\times A_2(\vec{k},\xi_{\vec{k}}+h\tilde{\nu}-\varepsilon).
	\end{multline}
We have pulled out a minus sign, because this term is negative: It corresponds
to a reduction of the thermal population in the final state as given by
Eq.~(\ref{eq:nk0}), induced by transitions to the initial state. These terms
describe an inverse rf spectroscopy analogous to the inverse photoemission in
condensed-matter systems. In the usual experimental practice, they do not
contribute because the atom cloud is prepared in a slightly out-of-equilibrium
state, where level $|3\rangle$ is empty. The resolution function in
Eq.~(\ref{eq:nkIb}) is
	\begin{multline}\label{eq:RIb}
		\mathscr{R}_{\vec{k}}^{\text{(Ib)}}(\varepsilon)=\frac{2}{\hbar^2}
		\int_{-\infty}^{\infty} d\varepsilon'd\varepsilon''
		\frac{f(\xi_{\vec{k}}+h\tilde{\nu}-\varepsilon')}{f(\xi_{\vec{k}}+h\nu_0)}\\
		\times A_3(\vec{k},\xi_{\vec{k}}+h\tilde{\nu}-\varepsilon')\,
		A_3(\vec{k},\xi_{\vec{k}}+h\tilde{\nu}-\varepsilon'')\\
		\times\text{Re}\,\int_{-\infty}^t\!\!dt'\,
		e^{i\frac{\varepsilon-\varepsilon''}{\hbar}(t-t')}\mathscr{E}(t')
		\int_{-\infty}^{t'}\!\!dt''\,e^{-i\frac{\varepsilon-\varepsilon'}{\hbar}(t-t'')}
		\mathscr{E}(t'').
	\end{multline}
	\end{subequations}
For a noninteracting final state, $\varepsilon'$ and $\varepsilon''$ are both
equal to $-h\nu$. The upper limit of the $t''$ integral can be extended from
$t'$ to $t$, correcting with a factor $1/2$. This shows that Eq.~(\ref{eq:RIb})
reduces to Eq.~(\ref{eq:RIp}) and that the two resolution functions are equal
for a noninteracting final state. We finally note that, if $t=+\infty$---i.e.,
if the measurement of the momentum distribution is performed after the
extinction of the rf pulse---the resolution functions are simply given by
	\begin{equation}
		\mathscr{R}^{\text{(I$'$)}}(\varepsilon)
		=\frac{1}{\hbar^2}\left|\mathscr{E}(\varepsilon/\hbar+2\pi\nu)\right|^2,
	\end{equation}
with $\mathscr{E}(\omega)$ the Fourier transform of $\mathscr{E}(t)$.

\section{Noninteracting final state}
\label{sec:no-interaction}

In this section, we neglect the interaction between the final state $|3\rangle$
and states $|1\rangle$ and $|2\rangle$ ($V_{13}=V_{23}=V_{31}=V_{32}=0$). We
furthermore restrict to a short-range interaction such that
$V_{\alpha\alpha}=0$. The atoms are free fermions in the final state, and the
nonzero matrix elements are $V_{12}=V_{21}$, describing the short-range
interaction between the states $|1\rangle$ and $|2\rangle$. In this limit,
diagram (I$'$) in Fig.~\ref{fig:fig03} gives the whole second-order response,
and the momentum distribution is the sum of Eqs.~(\ref{eq:nk00}),
(\ref{eq:nkIa}), and (\ref{eq:nkIb}). Because the spectral function in the final
state is a $\delta$ function, both resolution functions are given by
Eq.~(\ref{eq:RIp}). In the context of electron photoemission, an analogous model
known as the ``sudden approximation'' assumes a free-electron final state. In
contrast to rf spectroscopy for cold atoms, however, this remains an
approximation even in the ideal situation of a truly noninteracting final state,
because other effects (in particular the surface) are usually neglected as well.

\begin{figure}[tb]
\includegraphics[width=0.8\columnwidth]{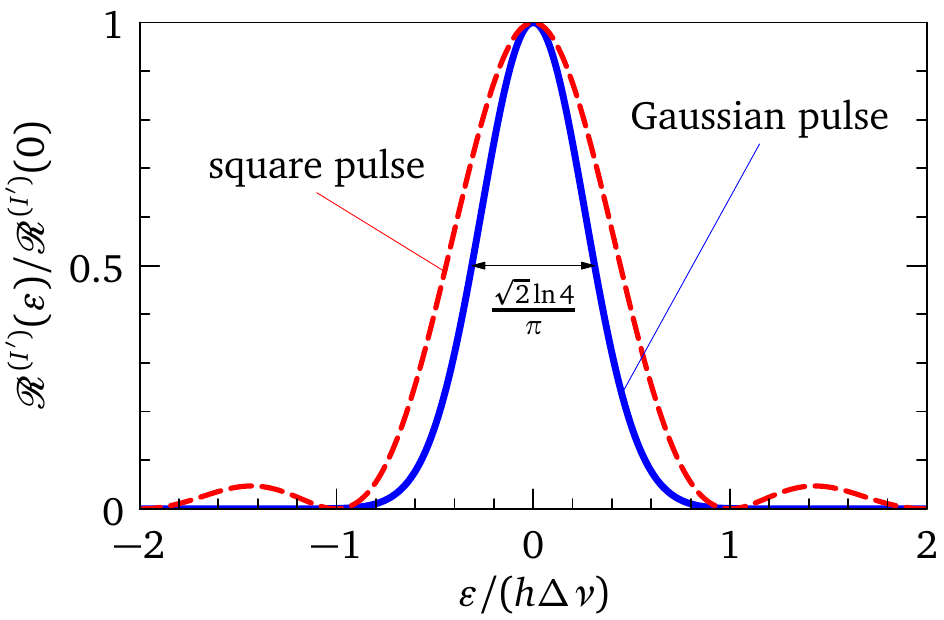}
\caption{\label{fig:fig04}(Color online)
Resolution function for a rf pulse with a square (dashed line) and Gaussian
(solid line) envelope in the case of a noninteracting final state. The width of
the envelope in the time domain is fixed by $\Delta\nu^{-1}$ in both cases (full
width at half maximum in the Gaussian case). The curves are normalized to the
peak maximum for easier comparison.
}
\end{figure}

We consider a monochromatic radiation of frequency $\nu$ with a slowly varying
envelope, such that the coupling in Eq.~(\ref{eq:conversion}) is
	$
		\mathscr{E}(t)=\bar{\mathscr{E}}(t)\cos(2\pi\nu t).
	$
Assuming that the momentum distribution is measured after the end of the pulse,
and that the duration of the pulse is much longer than $1/\nu$, the resolution
function is
	$
		\mathscr{R}^{\text{(I$'$)}}(\varepsilon)=
		|\bar{\mathscr{E}}(\varepsilon/\hbar)|^2/(4\hbar^2)
	$
with $\bar{\mathscr{E}}(\omega)$ the Fourier transform of
$\bar{\mathscr{E}}(t)$. For a square pulse of intensity $\mathscr{E}_0$ and
duration $\Delta\nu^{-1}$ in the limit $\nu\gg\Delta\nu$, the resolution
function is
	\begin{equation}\label{eq:Rsquare}
		\mathscr{R}^{\text{(I$'$)}}(\varepsilon)=
		\left(\frac{\pi\mathscr{E}_0}{h\Delta\nu}\right)^2
		\left[\frac{\sin(\pi x)}{\pi x}\right]^2,\quad
		x=\frac{\varepsilon}{h\Delta\nu}.
	\end{equation}
For a Gaussian pulse of the same intensity at the maximum and a full width at
half maximum $\Delta\nu^{-1}$ we have
	\begin{equation}\label{eq:Rgauss}
		\mathscr{R}^{\text{(I$'$)}}(\varepsilon)
		=\left(\frac{\pi\mathscr{E}_0}{h\Delta\nu}\right)^2
		\frac{\pi}{4\ln 2}\exp\left[-\frac{(\pi x)^2}{\ln 4}\right].
	\end{equation}
The two functions are compared in Fig.~\ref{fig:fig04}. The validity of any
approach based on equilibrium response is limited to a regime where the fraction
of atoms transferred to the final state is small, or equivalently, the time $t$
of the measurement is short compared with $2h/\mathscr{E}_0$, which is the
period of Rabi oscillations between the states $|2\rangle$ and $|3\rangle$.
Since, on the other hand, $t\gtrsim\Delta\nu^{-1}$, this regime corresponds to
$\mathscr{E}_0/(2h\Delta\nu)\ll1$. Our formalism is therefore valid as long as
the amplitude of the resolution function is much smaller than unity.

\subsection{Free and nearly free fermions in a harmonic trap}

If all interactions are turned off, the spectral function is
$A_2(\vec{k},\varepsilon)=\delta(\varepsilon-\xi_{\vec{k}})$. For a homogeneous
fermion gas, the momentum distribution is therefore simply
	\begin{multline}\label{eq:nkt0}
		\langle n_{\vec{k}}\rangle=\mathscr{R}^{\text{(I$'$)}}(h\tilde{\nu})f(\xi_{\vec{k}})\\
		+[1-\mathscr{R}^{\text{(I$'$)}}(h\tilde{\nu})]f(\xi_{\vec{k}}+h\nu_0).
	\end{multline}
The first term corresponds to the atoms excited from the initial state, while
the second term corresponds to the equilibrium thermal population of the final
state, reduced by the transitions to the initial state. From here on, we assume
that $h\nu_0$ is large enough for the second term in Eq.~(\ref{eq:nkt0}) to be
negligible. Equation~(\ref{eq:nkt0}) indicates that one can, in principle,
determine the frequency $\nu_0$ of the noninteracting transition, the resolution
$\Delta\nu$, and the temperature $T$ by measuring the momentum distribution with
all interactions suppressed: the energy-distribution curve (EDC) is just the
resolution function, while the momentum-distribution curve (MDC) is controlled
by the Fermi function. Experiments with homogeneous Fermi gases have not been
conducted yet (for bosons, see Ref.~\onlinecite{Gaunt-2013}). In this section,
we study within LDA the modifications of Eq.~(\ref{eq:nkt0}) due to the
nonhomogeneous distribution of atoms trapped in a harmonic potential, in two and
three dimensions. The resulting equations provide a way of determining the total
number of atoms, in addition to $\nu_0$, $\Delta\nu$, and $T$, by fitting
experimental EDCs and MDCs.

Consider a harmonic trap described by the potential $V(r)=(1/2)m\omega_r^2r^2$.
In dimension $d$, the number of atoms in state $|1\rangle$ is related to the
chemical potential by
	\begin{equation}\label{eq:N2-nointeraction}
		N_1=\int d^dr\int\frac{d^dk}{(2\pi)^d}\frac{1}{e^{\beta(\varepsilon_{\vec{k}}-\mu+
		\frac{1}{2}m\omega_r^2r^2)}+1}.
	\end{equation}
For free particles with a dispersion $\varepsilon_{\vec{k}}=\hbar^2k^2/(2m)$,
the evaluation of the integrals gives
	\begin{equation}\label{eq:N2-nointeraction1}
		N_1=-\left(\frac{k_{\text{B}}T}{\hbar\omega_r}\right)^d\text{Li}_d
		\left(-e^{\frac{\mu}{k_{\text{B}}T}}\right),
	\end{equation}
where $\text{Li}_2$ and $\text{Li}_3$ are the di- and trilogarithm,
respectively. Note that, unlike Eq.~(\ref{eq:N2-nointeraction1}) suggests, $\mu$
does depend on the particle mass $m$, because $\omega_r=(\kappa/m)^{1/2}$, where
$\kappa$ is the strength of the harmonic potential. To estimate the
trap-averaged momentum distribution, we replace $f(\xi_{\vec{k}})$ in
Eq.~(\ref{eq:nkt0}) with $f\big(\varepsilon_{\vec{k}}-\mu+V(r)\big)$, and we
perform a spatial integration. The result is
	\begin{equation}\label{eq:nkt0LDA}
		\langle n_{\vec{k}}\rangle_{\text{LDA}}=-\left(\frac{2\pi k_{\text{B}}T}
		{m\omega_r^2}\right)^{\frac{d}{2}}\text{Li}_{\frac{d}{2}}
		\left(-e^{\frac{\mu-\varepsilon_{\vec{k}}}{k_{\text{B}}T}}\right)
		\mathscr{R}^{\text{(I$'$)}}(h\tilde{\nu}).
	\end{equation}
Note that $\langle n_{\vec{k}}\rangle_{\text{LDA}}$ is extensive and has the
units of a normalization volume. Equation~(\ref{eq:nkt0LDA}) shows that for free
particles in the LDA, the EDCs are not affected by the inhomogeneities, because
the latter do not change the energy of the $|2\rangle\to|3\rangle$ transition.
The measured EDC line shape depends neither on the details of the density
distribution in the trap nor on the momentum, but is entirely determined by the
properties of the rf pulse.

A qualitative understanding of the effects of interactions on the EDC and MDC
curves may be gained by considering nearly free fermions. The simplest model is
that of free fermions with an effective mass $m^*$. With the caveat that such a
model can only be envisioned as a low-energy idealization, this effective mass
can be simulated by assuming for the bare fermions a self-energy,
	\begin{equation}\label{eq:Sigma-mstar}
		\Sigma_{\vec{k}}=\frac{\hbar^2k^2}{2}\left(\frac{1}{m^*}-\frac{1}{m}\right),
	\end{equation}
such that $\varepsilon_{\vec{k}}+\Sigma_{\vec{k}}=\hbar^2k^2/(2m^*)\equiv
E_{\vec{k}}$, and the spectral function of the initial state is
$A_2(\vec{k},\varepsilon)= \delta(\varepsilon-\xi_{\vec{k}}-\Sigma_{\vec{k}})$.
Neglecting the population of the final state, the corresponding momentum
distribution for a homogeneous gas is
	\begin{equation}
		\langle n_{\vec{k}}\rangle=\mathscr{R}^{\text{(I$'$)}}(h\tilde{\nu}-\Sigma_{\vec{k}})
		f(E_{\vec{k}}-\mu).
	\end{equation}
The maximum of the EDC is at $h(\nu_0-\nu)=\Sigma_{\vec{k}}$, and tracks the
difference in the dispersions of the initial and final states. The dispersion of
the EDC maximum is given by $E_{\text{max}}(k)=(\hbar^2k^2/2)(1/m^*-1/m)$. This
means that, for $m^*>m$, the peak moves towards lower values of the detuning
$\nu_0-\nu$ with increasing momentum $k$.

Like for free fermions, the inhomogeneities due to trapping do not affect the
EDC line shape for nearly free fermions, because the self-energy
(\ref{eq:Sigma-mstar}) does not depend on the local atom density. In such a gas,
a plot of the quadratic EDC peak dispersion as a function of $k$ gives the
effective mass. The MDC profile also reflects the effective mass.
Equation~(\ref{eq:N2-nointeraction1}) gets corrected by a factor $(m^*/m)^{d/2}$
because $\omega_r$ is defined in terms of the bare mass; in
Eq.~(\ref{eq:nkt0LDA}), the changes $\varepsilon_{\vec{k}}\to E_{\vec{k}}$ and
$h\tilde{\nu}\to h\tilde{\nu}-\Sigma_{\vec{k}}$ must be made in order to
describe harmonically trapped nearly free fermions.

\subsection{EDC dispersion due to inhomogeneity and Hartree shifts}
\label{sec:mH}

For free and nearly free fermions, the EDC line shape is not modified by the
inhomogeneity, and  the dispersion of the EDC peak tracks the intrinsic
quasiparticle dispersion. However, if the self-energy depends on density, these
convenient properties are lost. In order to illustrate this in the simplest
model, we consider the case of fermions subject to a short-range interaction,
which is treated to lowest order, by keeping only the Hartree term. The
momentum- and energy-independent Hartree self-energy in state $|2\rangle$ is
given by
	\begin{equation}\label{eq:Hartree_self-energy}
		\Sigma_2=(g/N_0)n_1,
	\end{equation}
with $n_1$ the density of atoms in state $|1\rangle$. The state $|1\rangle$
experiences a similar shift $\Sigma_1=(g/N_0)n_2$. The dimensionless coupling
$g$ is positive (negative) for repulsive (attractive) interaction, and $N_0$ is
the Fermi-level DOS, given by $mk_{\text{F}}/(2\pi^2\hbar^2)$ and
$m/(2\pi\hbar^2)$ in three and two dimensions, respectively. The coupling $g$ is
related	 to the scattering length via $g/N_0=4\pi\hbar^2a_{\text{3D}}/m$ and
$g=-1/\ln(k_{\text{F}}a_{\text{2D}})$ in 3D and 2D, respectively.
Equation~(\ref{eq:Hartree_self-energy}) means that the energy of the transition
is reduced (increased) with respect to the noninteracting value $h\nu_0$ for a
repulsive (attractive) interaction. In a harmonic trap, the modification varies
from the center to the periphery, and this contributes to a broadening and a
momentum dependence of the EDC, resulting in a dispersion of the EDC peak, as we
shall see. This dispersion may by qualified ``spurious'', because it is observed
in a system where the transition does not actually disperse with momentum.

\begin{figure}[tb]
\includegraphics[width=\columnwidth]{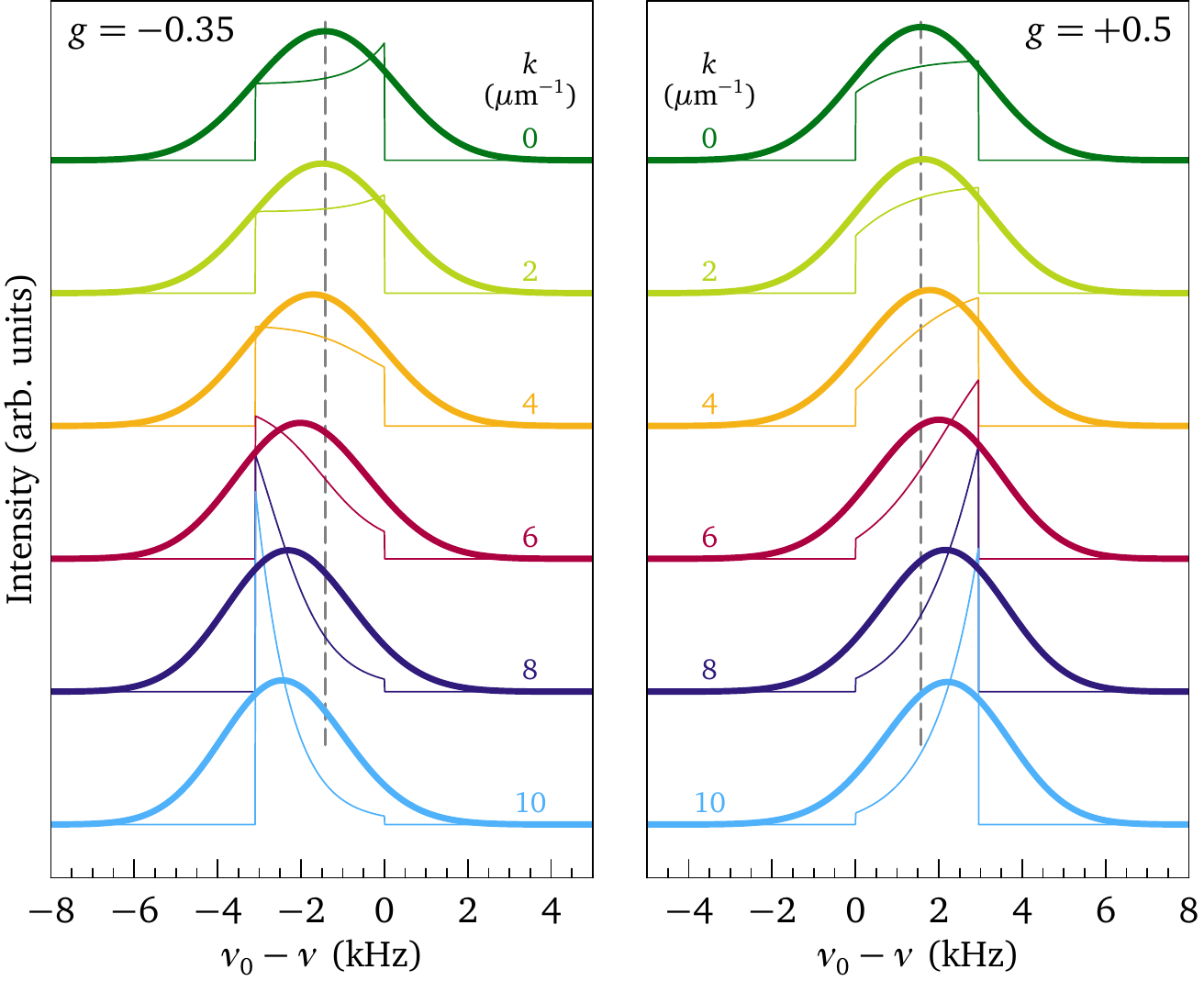}
\caption{\label{fig:fig05}(Color online)
Energy-distribution curves for harmonically trapped $^{40}$K atoms in two
dimensions: effect of the Hartree shift. The EDC (\ref{eq:nktgLDA}) for a
Gaussian pulse with $\Delta\nu=5$~kHz are normalized and shown at different
momenta (thick lines), for attractive (left) and repulsive (right) interaction.
The thin lines show the intrinsic EDC corresponding to each curve
[$\Delta\nu=0$, Eq.~(\ref{eq:nkgLDA0})], divided by two for clarity. The other
parameters are $\omega_r/2\pi=127$~Hz, $N_1=2000$, and $T=100$~nK.
}
\end{figure}

With the Hartree term (\ref{eq:Hartree_self-energy}), the spectral function is
$A_2(\vec{k},\varepsilon)=\delta(\varepsilon-\xi_{\vec{k}}-\Sigma_2)$. For a
homogeneous gas of density $n_1$, the chemical potential is set by the
self-consistency condition
	\begin{multline}
		n_1=\int\frac{d^dk}{(2\pi)^d}\int_{-\infty}^{\infty}
		d\varepsilon\,A_1(\vec{k},\varepsilon)f(\varepsilon)\\
		=-\left(\frac{mk_{\text{B}}T}{2\pi\hbar^2}\right)^{\frac{d}{2}}
		\text{Li}_{\frac{d}{2}}\left(-e^{\frac{\mu-(g/N_0)n_2}{k_{\text{B}}T}}\right),
	\end{multline}
and the momentum distribution (\ref{eq:nkIa}) becomes
	\begin{equation}
		\langle n_{\vec{k}}\rangle=\mathscr{R}^{\text{(I$'$)}}(h\tilde{\nu}-\Sigma_2)
		f(\xi_{\vec{k}}+\Sigma_2).
	\end{equation}
In a harmonic trap, the local self-consistency condition reads
	\begin{equation}\label{eq:self-consistent-Hartree}
		n_1(\vec{r})=-\left(\frac{mk_{\text{B}}T}{2\pi\hbar^2}\right)^{\frac{d}{2}}
		\text{Li}_{\frac{d}{2}}\left(
		-e^{\frac{\mu-\frac{1}{2}m\omega_r^2r^2-(g/N_0)n_2(\vec{r})}
		{k_{\text{B}}T}}\right),
	\end{equation}
where $\mu$ is fixed by the condition $N_1=\int d^dr\,n_1(\vec{r})$.
\footnote{The self-consistent equation (\ref{eq:self-consistent-Hartree}) breaks
down for $g\leqslant-1$. In this regime of interaction, the negative pressure
due to the Hartree term is stronger than the pressure due to Pauli exclusion, so
that $d\mu/dn<0$. In this case, the density profile implied by
Eq.~(\ref{eq:self-consistent-Hartree}) has a minimum at the center of the trap.}
The explicit expression for the momentum distribution in the harmonic trap is
therefore
	\begin{equation}\label{eq:nktgLDA}
		\langle n_{\vec{k}}\rangle_{\text{LDA}}=\int d^dr\,
		\frac{\mathscr{R}^{\text{(I$'$)}}
		\big(h\tilde{\nu}-(g/N_0)n_1(\vec{r})\big)}
		{\exp\left(\frac{\varepsilon_{\vec{k}}-\mu+\frac{1}{2}m\omega_r^2r^2
		+(g/N_0)n_1(\vec{r})}{k_{\text{B}}T}\right)+1}.
	\end{equation}
Figure~\ref{fig:fig05} shows the EDCs calculated using Eq.~(\ref{eq:nktgLDA}),
in a two-dimensional gas of $^{40}$K atoms with $n_1=n_2\equiv n$, and using
parameters typical for the experiment of Ref.~\onlinecite{Frohlich-2012}. The
maximum of the EDC disperses towards lower (higher) values of $\nu_0-\nu$ for
attractive (repulsive) interaction. The width of the EDC varies with momentum
and is larger than the expected resolution, which is
$\Delta\nu\sqrt{2}\ln4/\pi=3.12$~kHz.

\begin{figure}[tb]
\includegraphics[width=0.7\columnwidth]{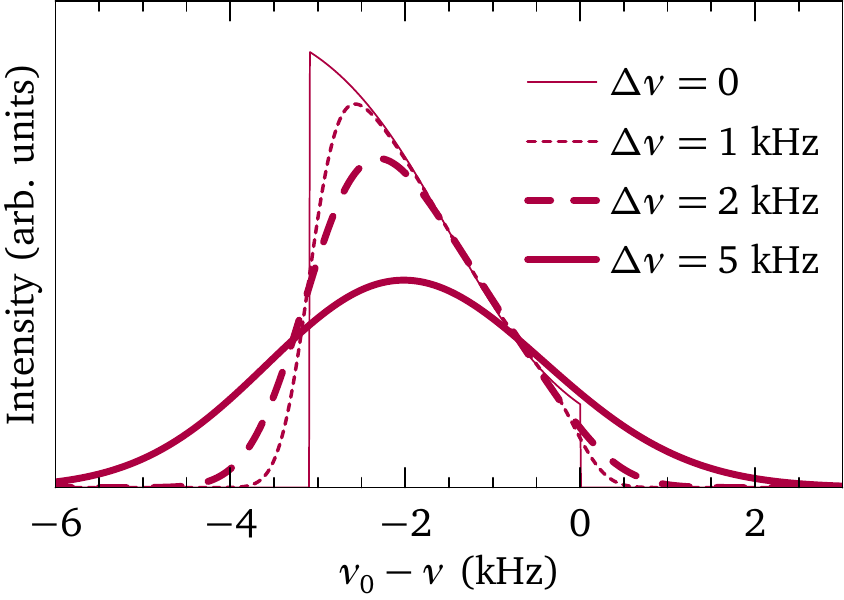}
\caption{\label{fig:fig06}(Color online)
Convergence of the measured EDC towards the intrinsic EDC (thin line) with
improving the resolution by increasing the duration $\Delta\nu^{-1}$ of the
Gaussian pulse. The EDCs are shown for $g=-0.35$ and $k=6~\mu\text{m}^{-1}$,
with the same parameters as in Fig.~\ref{fig:fig05}.
}
\end{figure}

\begin{figure}[tb]
\includegraphics[width=\columnwidth]{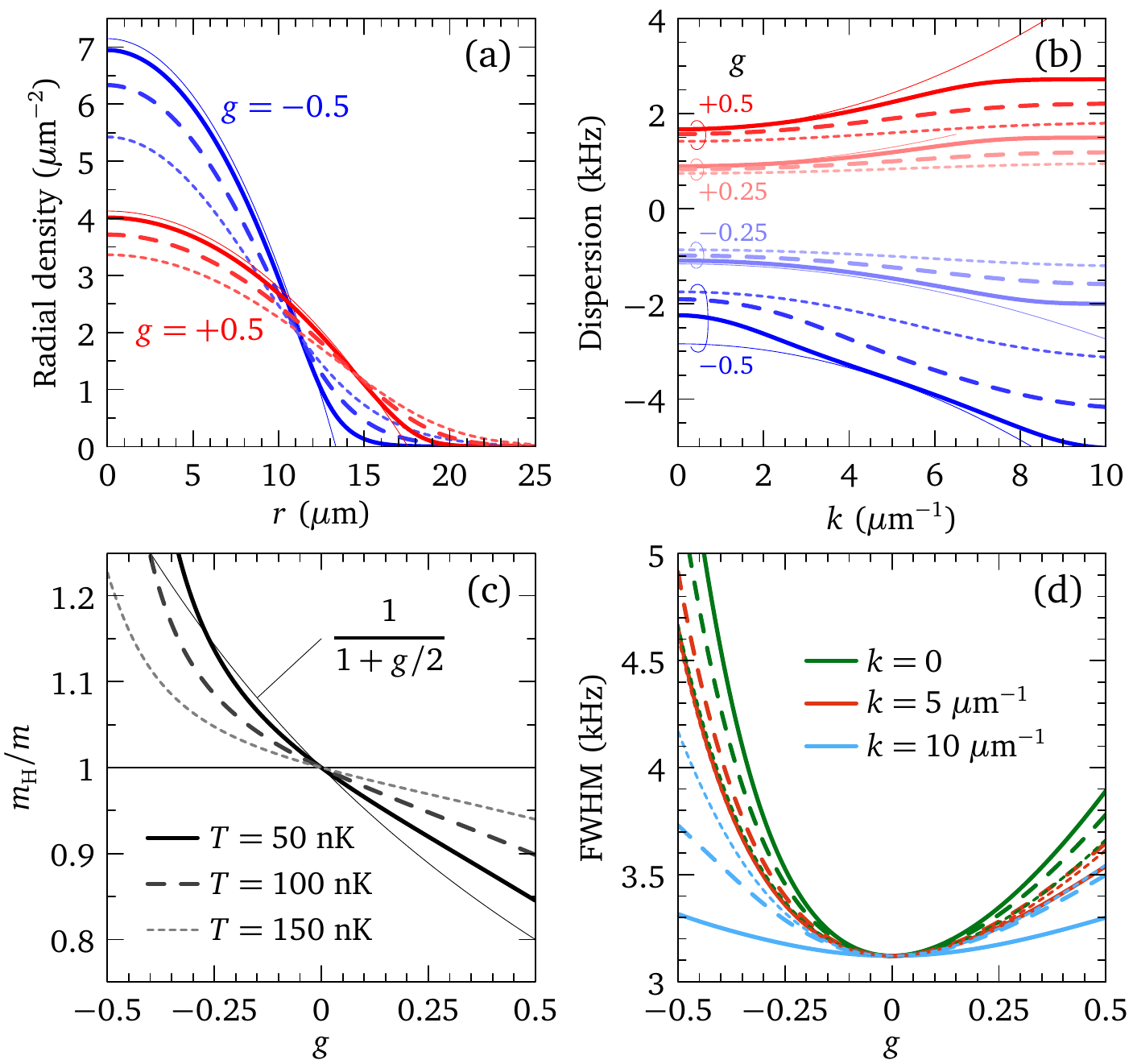}
\caption{\label{fig:fig07}(Color online)
(a) Radial density, (b) dispersion of the EDC maximum as a function of momentum,
(c) Hartree ``effective mass'' as a function of $g$, and (d) width of the EDC.
In all graphs, dotted lines correspond to $T=150$~nK, dashed lines to
$T=100$~nK, thick solid lines to $T=50$~nK, and the thin solid lines give the
analytical result for $T=0$ (see text). In (d), the width is shown as a function
of $g$ for three momenta; the colors correspond to those used in
Fig.~\ref{fig:fig05}. The model parameters are $\omega_r/2\pi=127$~Hz,
$N_1=2000$, and $\Delta\nu=5$~kHz. The minimum at $g=0$ in (d) corresponds to
the resolution of 3.12~kHz.
}
\end{figure}

The curves for attractive and repulsive interaction look similar in
Fig.~\ref{fig:fig05}; however, the magnitudes of $g$ are different. In fact,
there is a systematic asymmetry between positive and negative $g$, because an
attractive interaction tends to gather atoms near the center of the trap,
leading to a more inhomogeneous density [see Fig.~\ref{fig:fig07}(a)]. The width
of the EDC reflects the distribution of densities in the trap. This distribution
is defined as $D(\mathfrak{n})=\int
d^dr\,\delta\big(\mathfrak{n}-n(\vec{r})\big)$, and takes nonzero values for
densities $\mathfrak{n}$ between $0$ and $n(0)$. As shown in
Appendix~\ref{app:nkD}, it is possible to rewrite the momentum distribution
(\ref{eq:nktgLDA}) as an integral over densities involving $D(\mathfrak{n})$
[Eq.~(\ref{eq:nkgLDA_D})]. In dimension $d=2$, the distribution
$D(\mathfrak{n})$ can be evaluated explicitly (see Appendix~\ref{app:nkD}). For
an ideal resolution, the resulting momentum distribution is
	\begin{equation}\label{eq:nkgLDA0}
		\langle n_{\vec{k}}\rangle_{\text{LDA}}\propto\begin{cases}
		\displaystyle\frac{1+g+b\left(\frac{h\tilde{\nu}}{g}\right)}
		{1+e^{\beta\varepsilon_k}b\left(\frac{h\tilde{\nu}}{g}\right)}
		& 0\leqslant\frac{h\tilde{\nu}}{g}\leqslant\frac{n(0)}{N_0}\vspace{2mm}\\
		0 & \text{otherwise,}\end{cases}
	\end{equation}
with $b(\varepsilon)=(e^{\beta\varepsilon}-1)^{-1}$. This is shown as the thin
lines in Fig.~\ref{fig:fig05} and corresponds to the $\Delta\nu\to0$ limit of
Eq.~(\ref{eq:nktgLDA}). The peculiar line shape (\ref{eq:nkgLDA0}), which
depends on both the density distribution $D(\mathfrak{n})$ and the Fermi
occupation factors, could be revealed experimentally by a moderate improvement
of the resolution. Figure~\ref{fig:fig06} shows the evolution of a typical line
shape, as the full width at half maximum of the Gaussian rf pulse is increased
from 0.2 to 1.0~ms.

The dispersion of the EDC maximum with increasing momentum is plotted in
Fig.~\ref{fig:fig07}(b) for various interaction strengths and temperatures. This
``spurious'' dispersion may be understood as follows. At each point in the trap,
the minimum of the local energy band is the sum of the harmonic potential and
the Hartree term. The $k=0$ states are occupied throughout the trap and give
contributions to the EDC with a Hartree shift ranging between 0 at the periphery
and $(g/N_0)n(0)$ at the center. The EDC for $k=0$ extends therefore from
$\tilde{\nu}=0$ to $\tilde{\nu}=(g/N_0)n(0)/h$. As the momentum increases, the
corresponding $k$ states in the low-density regions at the periphery are above
the chemical potential, and their thermal population contributes less to the
intrinsic EDC. The latter is depressed near $\tilde{\nu}=0$ and becomes
asymmetric. Once filtered with the resolution function, the observed EDC
disperses as seen in Fig.~\ref{fig:fig05}. We can be more quantitative in the
limit $T\to0$, where the intrinsic EDC (\ref{eq:nkgLDA0}) becomes a rectangular
distribution, constant for $\varepsilon_k\leqslant h\tilde{\nu}/g\leqslant
n(0)/N_0$, and zero otherwise. Convolved with the resolution function, this
distribution gives a peak whose maximum disperses quadratically:
$E_{\text{max}}(k)=(g/2)[\varepsilon_k+n(0)/N_0]$. This dispersion can be
parametrized by a Hartree ``effective mass'' $m_{\text{H}}$ as
$E_{\text{max}}(k)-E_{\text{max}}(0)=(\hbar^2k^2/2)\left(1/m_{\text{H}}-
1/m\right)$. We then find $m_{\text{H}}/m=1/(1+g/2)$. This is compared in
Fig.~\ref{fig:fig07}(c) with the mass calculated numerically for various
temperatures. The density at the trap center can also be evaluated at $T=0$:
$n(0)/N_0=\hbar\omega_r\sqrt{2N_1/(1+g)}$.\cite{Note1} With this, we can
calculate the full $T=0$ dispersion, which is shown as thin lines in
Fig.~\ref{fig:fig07}(b).

At finite $T$, the dispersion is not quadratic, except close to $k=0$, and for
$g$ not too close to $-1$. As temperature increases, the particle cloud spreads
more across the trap [see Fig.~\ref{fig:fig07}(a)], the density distribution
$D(\mathfrak{n})$ sharpens, and the peak dispersion therefore gets weaker. The
asymmetry between repulsive and attractive interaction is strongest at $T=0$ and
is reduced as temperature increases. The finite-$T$ Hartree mass $m_{\text{H}}$,
deduced from the curvature of the dispersion at $k=0$, is shown in
Fig.~\ref{fig:fig07}(c). The relative mass is larger than unity for attractive
interaction and smaller than unity for repulsive interaction. Its dependence on
temperature is linear for $g>0$, but more complicated for $g<0$; in particular,
nonlinearities in the temperature dependence get stronger as $g$ approaches
$-1$, as can be seen in Figs.~\ref{fig:fig07}(b) and \ref{fig:fig07}(c).

As seen in Fig.~\ref{fig:fig05}, the EDC not only disperses due to
inhomogeneity, but also narrows with increasing momentum. At zero temperature,
the width of the EDC has a complicated dependency on $\Delta\nu$, which
approaches a linear function of $|g|$ as $\Delta\nu\to0$, namely
$|g/h(\varepsilon_k-n(0)/N_0)|$. At finite temperature and finite resolution,
however, the EDC width behaves more like $\sim g^2$, as shown in
Fig.~\ref{fig:fig07}(d). At $k=0$, the width reflects the radial density
distribution of Fig.~\ref{fig:fig07}(a): It is larger for attractive than for
repulsive interaction of the same magnitudes and decreases with increasing
temperature. The rounded behavior at $g=0$ transforms into a linear behavior
$\sim|g|/\sqrt{1+g}$ as $\Delta\nu$ is reduced.\cite{Note1} At large momenta,
the width is controlled by the Fermi edge rather than by $D(\mathfrak{n})$: It
is resolution-limited at low temperature and increases with increasing $T$.

The dispersion displayed in Figs.~\ref{fig:fig05} and \ref{fig:fig07}(b), as
well as a width looking like $\sim g^2$ as seen in Fig.~\ref{fig:fig07}(d),
could easily be mistaken as a signature of dynamical interactions, since this is
the expected behavior in a homogeneous Fermi liquid. Even the narrowing of the
EDC with increasing $k$ could evoke the sharpening of quasiparticles when
approaching the Fermi momentum. One feature, however, among these
inhomogeneity-driven effects is contrary to the expected signature of
interactions: The sharpening of the $k=0$ EDC with increasing $T$, due to the
flattening of the atom cloud in the trap, cannot be reconciled with the expected
increase of the scattering rate with $T$ in an interacting system.

\begin{figure}[tb]
\includegraphics[width=\columnwidth]{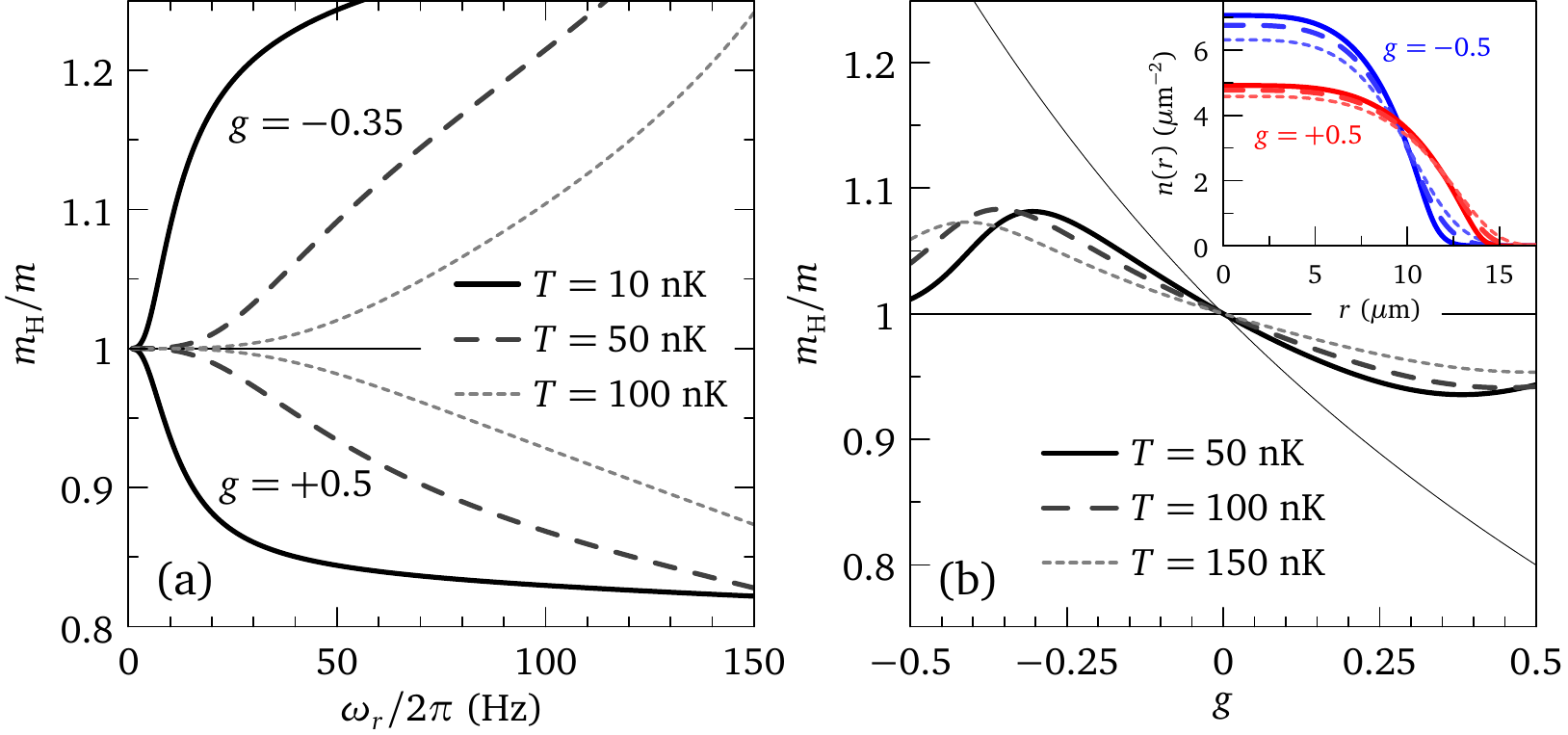}
\caption{\label{fig:fig08}(Color online)
(a) Evolution of the Hartree ``effective mass'' with $\omega_r$ for attractive
(top curves) and repulsive (bottom curves) interaction and for three different
temperatures. (b) Hartree ``effective mass'' as a function of $g$ for a $r^4$
trap and three temperatures. The thin solid line shows $1/(1+g/2)$ for easier
comparison with Fig.~\ref{fig:fig07}(c). (Inset) Radial density for two values
of $g$ and the same three temperatures.
}
\end{figure}

An obvious way to reduce the spurious mass $m_{\text{H}}$ in experiments is to
achieve a more homogeneous density. In three dimensions, this can be realized by
means of a weaker confining harmonic potential. It does not work well in two
dimensions, though, because the density distribution $D(\mathfrak{n})$ is flat
at zero temperature: All densities between zero and the maximum density are
equally represented, irrespective of the potential strength. This is illustrated
by the fact that $m_{\text{H}}$ is given by $1/(1+g/2)$ at zero temperature,
which does not depend on $\omega_r$. At finite $T$, $D(\mathfrak{n})$ does
depend on $\mathfrak{n}$ (see Appendix~\ref{app:nkD}), and thus $m_{\text{H}}/m$
approaches unity as $\omega_r$ is reduced, as shown in Fig.~\ref{fig:fig08}(a).
Alternatively, one may obtain a more homogeneous density by means of a quartic
trapping potential. Figure~\ref{fig:fig08}(b) shows $m_{\text{H}}$ as a function
of $g$ when the harmonic trap is replaced by a $r^4$ trap. The latter was
defined such that the potential equals the harmonic potential with
$\omega_r/2\pi=127$~Hz at a distance $r=10~\mu$m. The inset shows that the
density profile is flatter than in Fig.~\ref{fig:fig07}(a), and as a result
$m_{\text{H}}$ is significantly reduced as compared to Fig.~\ref{fig:fig07}(c).
The spurious dispersion can also be reduced, and in some cases even suppressed,
by taking advantage of interactions in the final state; this is discussed in
Sec.~\ref{sec:final-state2}.

\subsection{Inhomogeneity and momentum-integrated rf intensity}
\label{sec:Li6}

In this paragraph we briefly discuss the effect of Hartree shifts and
inhomogeneity on the momentum-integrated rf intensity. In the particular case of
a Gaussian density profile, which is a good approximation for three-dimensional
gases at not too low temperature, and a Gaussian rf pulse, the trap-averaged
integrated intensity takes a universal form depending on a single parameter
$\propto a_{\text{3D}}n(0)/(m\Delta\nu)$. We compare this form with the
measurements of Ref.~\onlinecite{Gupta-2003}, where Hartree shifts in $^6$Li
mixtures were studied by rf spectroscopy as a means to determine the scattering
length.

\begin{figure}[tb]
\includegraphics[width=0.9\columnwidth]{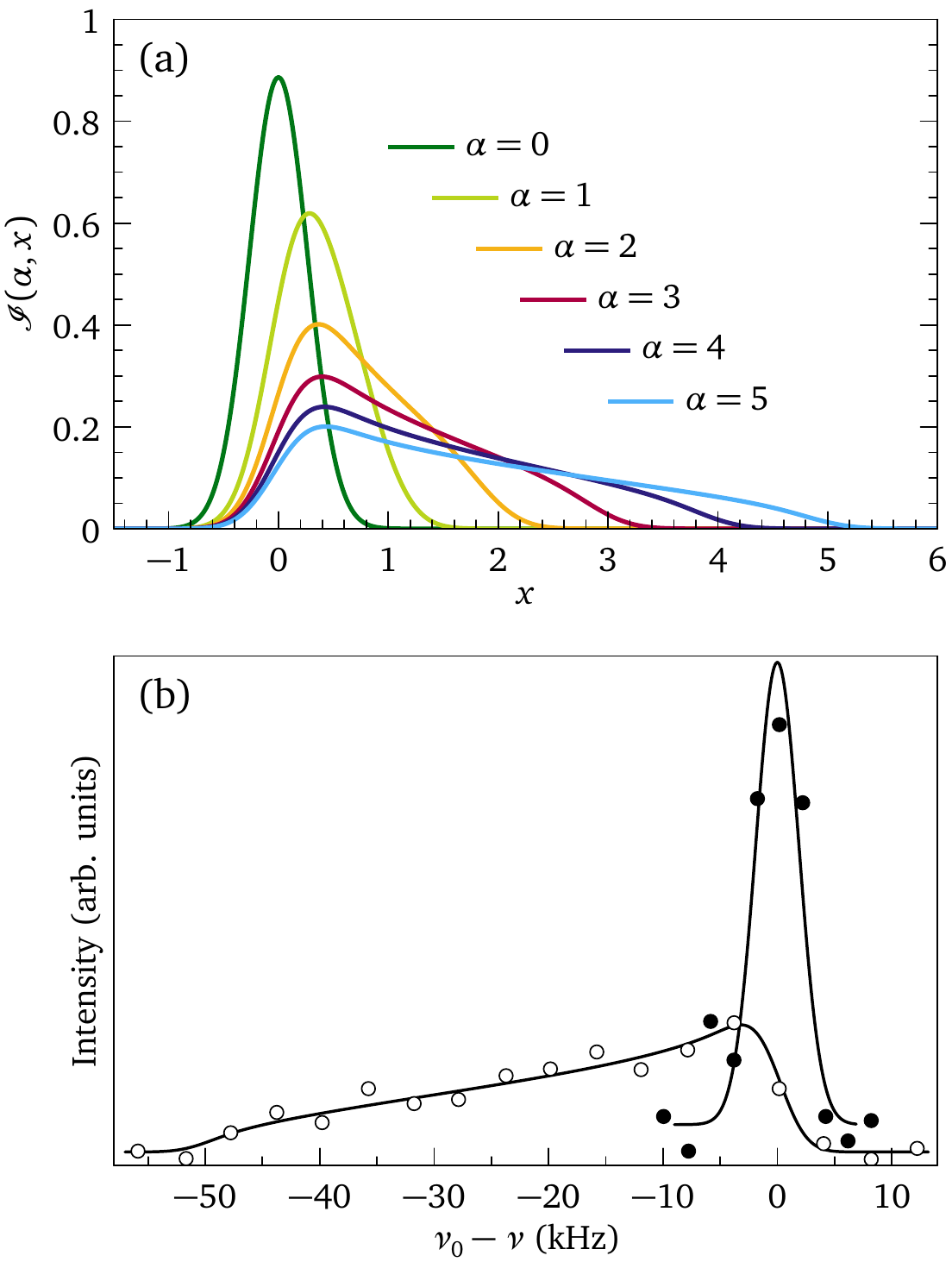}
\caption{\label{fig:fig09}(Color online)
(a) The function $\mathscr{I}(\alpha,x)$ defined in Eq.~(\ref{eq:functionI}) for
various values of $\alpha$. Note that the cutoff is at $x\approx\alpha$. (b)
Data of Ref.~\onlinecite{Gupta-2003} (dots) and fits to
Eq.~(\ref{eq:I_RF_LDA_3D}) with $\Delta\nu$ fixed to 7.1~kHz (lines). A constant
background was added to Eq.~(\ref{eq:I_RF_LDA_3D}) for fitting.
}
\end{figure}

The momentum integration of the rf intensity (\ref{eq:nktgLDA}) yields
	\begin{equation}
		I_{\text{LDA}}=\int d^dr\,n_2(\vec{r})\,\mathscr{R}^{\text{(I$'$)}}
		\big(h\tilde{\nu}-(g/N_0)n_1(\vec{r})\big).
	\end{equation}
For a balanced gas with $n_1=n_2=n$, this becomes a one-dimensional integral
involving the density distribution:
	\begin{equation}\label{eq:I_RF_LDA}
		I_{\text{LDA}}=\int_{-\infty}^{\infty}d\mathfrak{n}\,D(\mathfrak{n})\,\mathfrak{n}\,
		\mathscr{R}^{\text{(I$'$)}}\big(h\tilde{\nu}-(g/N_0)\mathfrak{n}\big).
	\end{equation}
We are interested in comparing this expression with the data of
Ref.~\onlinecite{Gupta-2003}, which were obtained at $T\sim0.7T_{\text{F}}$. At
such temperatures, the density profile given by
Eq.~(\ref{eq:self-consistent-Hartree}) is very close to a Gaussian in three
dimensions: $n(r)\approx
n(0)\exp\big\{-\pi\left[n(0)/N_1\right]^{2/3}r^2\big\}$. For this Gaussian
profile, the density distribution is
	\begin{equation}
		D(\mathfrak{n})=\begin{cases}
		\displaystyle\frac{2N_1\sqrt{\ln[n(0)/\mathfrak{n}]/\pi}}{n(0)\mathfrak{n}} &
		0\leqslant \mathfrak{n}\leqslant n(0)\vspace{2mm}\\
		0 &\text{otherwise}.\end{cases}
	\end{equation}
If the rf pulse has a Gaussian envelope, Eq.~(\ref{eq:I_RF_LDA}) becomes
	\begin{equation}\label{eq:I_RF_LDA_3D}
		I_{\text{LDA}}\propto\mathscr{I}\left(\frac{gn(0)}{h\Delta\nu N_0},
		\frac{\tilde{\nu}}{\Delta\nu}\right)
		=\mathscr{I}\left(\frac{2\hbar a_{\text{3D}}n(0)}{m\Delta\nu},
		\frac{\tilde{\nu}}{\Delta\nu}\right),
	\end{equation}
where the function $\mathscr{I}$ is given by
	\begin{equation}\label{eq:functionI}
		\mathscr{I}(\alpha,x)=\int_0^{\infty}du\,\sqrt{u}\,e^{-u}\,
		\exp\left[-\frac{\pi^2}{\ln4}(x-\alpha e^{-u})^2\right].
	\end{equation}
This function is displayed in Fig.~\ref{fig:fig09}(a). For $\alpha=0$, it is a
Gaussian, and for $\alpha\neq 0$ it has an asymmetric shape with a cutoff at
$x\approx\alpha$. Note that $\mathscr{I}(\alpha,x)=\mathscr{I}(-\alpha,-x)$, so
that symmetric curves are expected for attractive and repulsive interactions
corresponding to identical values of the product $|a_{\text{3D}}|n(0)$.

Equation (\ref{eq:I_RF_LDA_3D}) can be fit to the $^6$Li data as shown in
Fig.~\ref{fig:fig09}(b). The function $\mathscr{I}$ captures remarkably well the
peculiar asymmetry of the line shape in the presence of interactions (white
dots), in contrast to the Gaussian used in Ref.~\onlinecite{Gupta-2003},
resulting in an excellent fit. We have set $\Delta\nu=7.1$~kHz, which
corresponds to the Gaussian rf pulse of 140~$\mu$s used in this experiment. For
the data without interaction (black dots), there is no further adjustable
parameter, apart from the amplitude and a constant background. For the curve
with interaction, the fit yields $\alpha=-7.15\pm0.34$. Considering that the
average density is $\bar{n}\sim2.4\times10^{13}$~cm$^{-3}$, and that
$\bar{n}=n(0)/2^{3/2}$ for a Gaussian profile, this corresponds to a scattering
length $a_{\text{3D}}=-35.4\pm1.7$~nm. The analysis of
Ref.~\onlinecite{Gupta-2003} yields a value consistent within the error bars,
$-31\pm2.7$~nm, but we believe that Eq.~(\ref{eq:I_RF_LDA_3D}) provides a more
accurate way of measuring the scattering length.

\section{Discussion of final-state effects}
\label{sec:final-state}

\subsection{Resolution function for a Lorentzian final state}
\label{sec:final-state1}

\begin{figure}[tb]
\includegraphics[width=0.9\columnwidth]{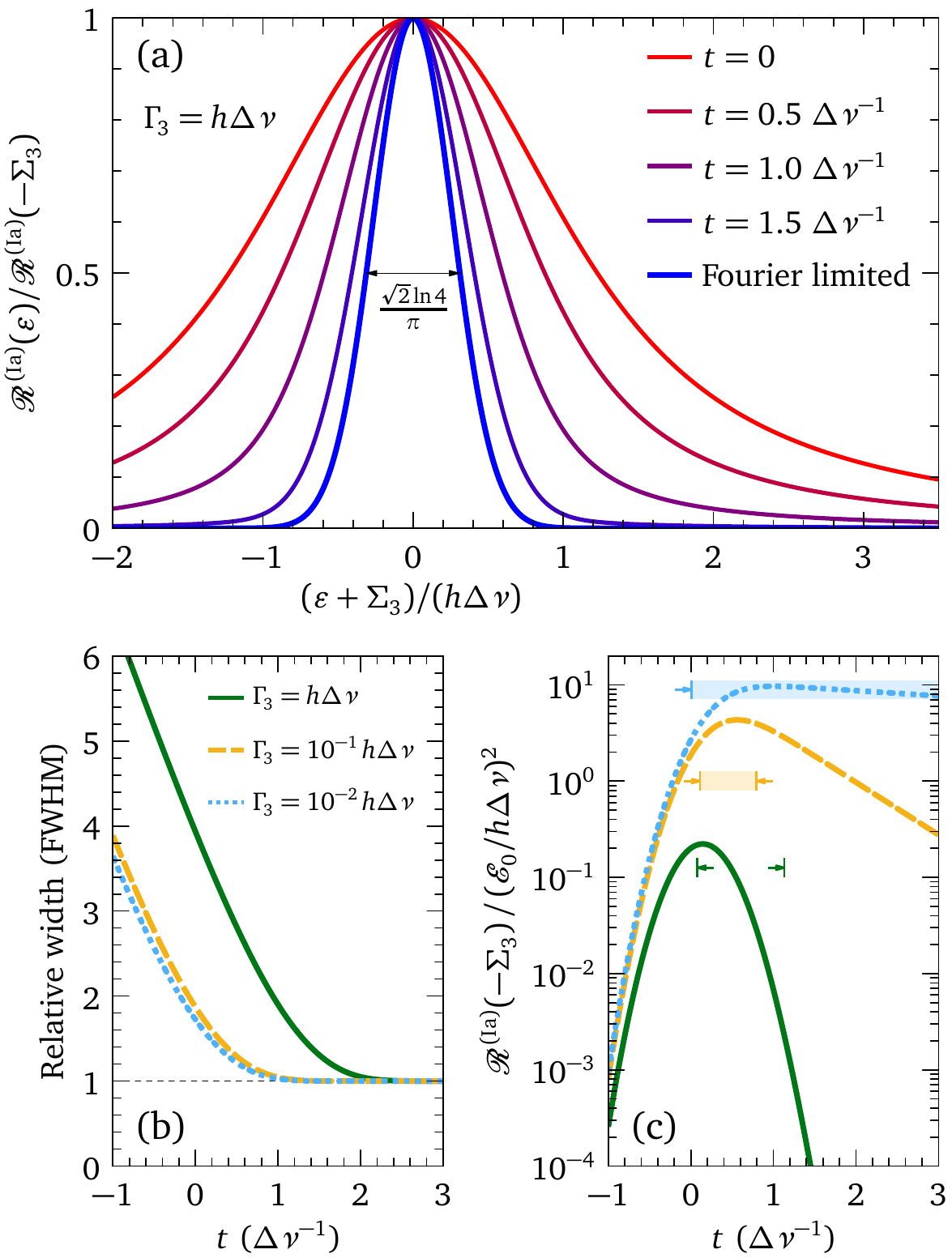}
\caption{\label{fig:fig10}(Color online)
(a) Energy dependence of the resolution function (\ref{eq:RIaLorentzian}) for a
final-state scattering rate $\Gamma_3=h\Delta\nu$, corresponding to a lifetime
$\tau_3=\Delta\nu^{-1}/(4\pi)$, and for increasing measurement times (broader to
narrower). The time $t$ is measured from the maximum of the Gaussian pulse
envelope, such that $t=\Delta\nu^{-1}$ corresponds to a measurement time one
full width after the pulse maximum. (b) Full width at half maximum of the
resolution function relative to the Fourier limited value and (c) maximum
intensity as a function of the measurement time and scattering rate. The arrows
pointing to the left in (c) indicate the time $\tau_3$, and those pointing to
the right the time $\frac{1}{16\ln 2\,\tau_3\Delta\nu}\Delta\nu^{-1}$.
}
\end{figure}

We start this section by deriving the modifications due to the resolution
functions (\ref{eq:RIp}) and (\ref{eq:Rgauss}), in the situation where
interactions lead to a shift and a lifetime for the final state of the rf
transition. We introduce these effects by means of a phenomenological
self-energy $\Sigma_3-i\Gamma_3$ in the final state, where $\Sigma_3$ is the
energy shift and $\Gamma_3$ is the scattering rate. These two quantities are, in
principle, related by causality and should be of the same order of magnitude for
weak interactions. The corresponding lifetime of the final state is
$\tau_3=\hbar/(2\Gamma_3)$, and the spectral function reads
	\begin{equation}
		A_3(\vec{k},\varepsilon)=\frac{\Gamma_3/\pi}
		{(\varepsilon-\xi_{\vec{k}}-h\nu_0-\Sigma_3)^2+\Gamma_3^2}.
	\end{equation}
The noninteracting result (\ref{eq:RIp}) gets modified like this:
	\begin{equation}
		\mathscr{R}^{\text{(Ia)}}(\varepsilon)=\frac{e^{-2\Gamma_3t/\hbar}}{\hbar^2}
		\scalebox{1}[1.3]{\Big|}
		\int_{-\infty}^tdt'\,e^{i(\varepsilon+h\nu+\Sigma_3-i\Gamma_3)t'/\hbar}\mathscr{E}(t')
		\scalebox{1}[1.3]{\Big|}^2.
	\end{equation}
The overall magnitude of the resolution function vanishes on time scales larger
than $\tau_3$, because atoms in the final state decay. Besides, the energy
dependence of the resolution function is also affected. In order to find out
how, we perform the time integration explicitly for the case of a Gaussian pulse
of full width at half maximum $\Delta\nu^{-1}$. In the relevant limit
$\Delta\nu\ll\nu$, the formula replacing Eq.~(\ref{eq:Rgauss}) is
	\begin{multline}\label{eq:RIaLorentzian}
		\mathscr{R}^{(\text{Ia})}(\varepsilon)=\left(\frac{\pi\mathscr{E}_0}{h\Delta\nu}\right)^2
		\frac{\pi}{16\ln 2}\exp\left[\frac{\pi^2}{\ln4}\left(\frac{\Gamma_3}{h\Delta\nu}\right)^2-
		\frac{2\Gamma_3t}{\hbar}\right]\\
		\times\exp\left[-\frac{\pi^2}{\ln4}\left(\frac{\varepsilon+\Sigma_3}{h\Delta\nu}\right)^2\right]\\
		\times\left|1+\text{erf}\left(2\sqrt{\ln2}\Delta\nu t-\frac{i\pi}{2\sqrt{\ln2}}
		\frac{\varepsilon+\Sigma_3-i\Gamma_3}{h\Delta\nu}\right)\right|^2.
	\end{multline}
This complicated expression has an interesting time dependence
(Fig.~\ref{fig:fig10}). The resolution function is even and centered at the
energy $\varepsilon=-\Sigma_3$, and it is significantly non-Gaussian when the
time delay $t$ of the measurement---counted in Eq.~(\ref{eq:RIaLorentzian}) from
the maximum of the pulse envelope---is comparable to the width of the pulse. For
large times $t\gg\Delta\nu^{-1}$, the erf function approaches one, and the
energy dependence of Eq.~(\ref{eq:RIaLorentzian}) measured from
$\varepsilon=-\Sigma_3$ is identical to the noninteracting result
(\ref{eq:Rgauss}), shown in Fig.~\ref{fig:fig10}(a) as ``Fourier limited''. The
width of the resolution function takes off for measurement times of the order of
$\Delta\nu^{-1}$ and increases roughly linearly with decreasing $t$
[Fig.~\ref{fig:fig10}(b)]. The peak intensity of
$\mathscr{R}^{(\text{Ia})}(\varepsilon)$ is largest shortly after the pulse
maximum and decreases for longer times [Fig.~\ref{fig:fig10}(c)].

Measurements can be done in the regime where the resolution function is Fourier
limited, provided that the time $t$ is smaller than the lifetime $\tau_3$, but
sufficiently large, that the real part of the argument in the erf function is
large and positive. These requirements read
	\begin{equation}
		\frac{1}{16\ln 2\,\tau_3\Delta\nu}\Delta\nu^{-1}<t<\tau_3.
	\end{equation}
Clearly, such a regime does not exist unless
$\tau_3\gg\Delta\nu^{-1}/(4\sqrt{\ln 2})$ or
$\Gamma_3\ll(\sqrt{\ln2}/\pi)h\Delta\nu$, as illustrated in
Fig.~\ref{fig:fig10}(c).

\subsection{Hartree shifts in the final state}
\label{sec:final-state2}

In Sec.~\ref{sec:mH}, we assumed $n_1=n_2$, which is justified if the fraction
of atoms transferred to the final state is small. A more accurate modeling of
experiments on balanced gases would be to take $n_2=n_1-n_3$, where $n_3$ is the
number of atoms in the final state. If $n_3$ is a fraction $f$ of $n_1$, we may
write $n_2=(1-f)n_1$ and $n_3=fn_1$. Let us furthermore take into account the
interactions $g_{13}$ and $g_{23}$ between states $|1\rangle$ and $|3\rangle$
and states $|2\rangle$ and $|3\rangle$, respectively, in addition to the
interaction $g_{12}$ (which was denoted $g$ in Sec.~\ref{sec:mH}). Treating all
interactions at first order, we find that the level $|\alpha\rangle$,
$\alpha=1,2,3$, is shifted by the self-energy
	\begin{equation}
		\Sigma_{\alpha}=(g^*_{\alpha}/N_0)n_1,
	\end{equation}
with $g^*_1=g_{12}(1-f)+g_{13}f$, $g^*_2=g_{12}+g_{23}f$, and
$g^*_3=g_{13}+g_{23}(1-f)$. The shift $\Sigma_3$ of the final state is larger at
the center of the trap than at the periphery and will therefore contribute to
the spurious mass $m_{\text{H}}$. We assume that $|\Sigma_3|$ remains much
smaller than $h\nu_0$, such that interaction effects related to the thermal
population of the final state are negligible.

The resolution function reflects the shift of the final state:
$\mathscr{R}^{\text{(Ia)}}_{\vec{k}}(\varepsilon) =
\mathscr{R}^{\text{(I$'$)}}(\varepsilon+\Sigma_3)$. As a result,
Eq.~(\ref{eq:nktgLDA}) is replaced with
	\begin{equation}
		\langle n_{\vec{k}}\rangle_{\text{LDA}}=\int d^dr\,
		\frac{\mathscr{R}^{\text{(I$'$)}}
		\left(h\tilde{\nu}-\frac{g^*_2-g^*_3}{N_0}n_1(\vec{r})\right)}
		{\exp\left(\frac{\varepsilon_{\vec{k}}-\mu+\frac{1}{2}m\omega_r^2r^2
		+(g^*_2/N_0)n_1(\vec{r})}{k_{\text{B}}T}\right)+1}.
	\end{equation}
One sees that the width of the rf signal is now controlled by $g^*_2-g^*_3$
instead of $g_{12}$. If the parameters (interactions and/or transferred fraction
$f$) can be arranged such that $g^*_2=g^*_3$, then the dispersion of the final
state locally follows the dispersion of the initial states, and no spurious
dispersion should be observed.

An explicit expression for the Hartree ``effective mass'' in the presence of
final-state shifts can be derived in two dimensions: The ideal
momentum-distribution line shape (\ref{eq:nkgLDA0}) is replaced with
	\begin{equation}
		\langle n_{\vec{k}}\rangle_{\text{LDA}}\propto\begin{cases}
		\displaystyle\frac{1+g^*_1+b\left(\frac{h\tilde{\nu}}{g^*_2-g^*_3}\right)}
		{1+e^{\beta\left(\varepsilon_k-\frac{g^*_1-g^*_2}{g^*_2-g^*_3}h\tilde{\nu}\right)}
		b\left(\frac{h\tilde{\nu}}{g^*_2-g^*_3}\right)}
		& 0\leqslant\frac{h\tilde{\nu}}{g^*_2-g^*_3}\leqslant\frac{n(0)}{N_0}\vspace{2mm}\\
		0 & \text{otherwise.}\end{cases}
	\end{equation}
In the limit $T\to0$, this becomes again a steplike distribution, whose center
disperses quadratically with momentum. Proceeding as in Sec.~\ref{sec:mH}, we
find
	\begin{equation}
		\frac{m_{\text{H}}}{m}=\frac{1+g^*_1-g^*_2}{1+g^*_1-(g^*_2+g^*_3)/2},
	\end{equation}
which is indeed unity if $g^*_2=g^*_3$.

\subsection{Vertex corrections at low temperature and density}
\label{sec:final-state3}

The vertex corrections of type II describe final-state effects going beyond the
self-energy renormalizations of the final state. We estimate such effects in
this section and indicate how they could be implemented to improve the
theoretical description of rf measurements. In the context of electron
photoemission, specific vertex corrections were shown to describe the production
of plasmons \cite{Chang-1973} or phonons \cite{Caroli-1973} during the
photoexcitation process. These phenomena are not relevant for cold-atom systems,
but other interesting effects take place, related to the spatial correlations
among the dilute atoms. We proceed in two steps, in order to identify the
important vertex diagrams. First, we consider the regime $k_{\text{B}}T\ll
h\nu_0$ and eliminate all diagrams that require a thermal population of the
final state. Then we organize the remaining diagrams according to the number of
hole lines in the initial states in the spirit of the low-density expansion for
the self-energy \cite{*[] [{ [Sov. Phys. JETP \textbf{7}, 104 (1958)].}]
Galitskii-1958}. We furthermore assume a short-range potential, such that the
interactions $V_{\alpha\alpha}$ are blocked by the Pauli principle.

\begin{figure}[tb]
\includegraphics[width=0.7\columnwidth]{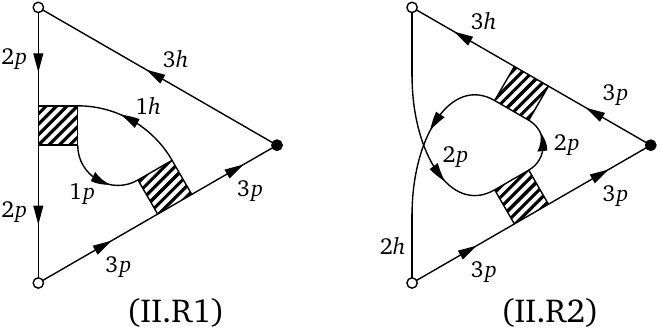}
\caption{\label{fig:fig11}
Dominant vertex corrections in the low-temperature low-density limit. The shaded
boxes represent a particle-particle ladder series (pseudopotential). $1p$ and
$1h$ stand for a particle or a hole in state $|1\rangle$, respectively, and
similarly for the other states. The diagrams give a significant contribution
only if the time ordering of the various vertices is such that all lines marked
as particles go to the right (see Appendix~\ref{app:type-II}). The diagrams
shown are right handed; there are two equivalent left-handed terms.
}
\end{figure}

This analysis, outlined in Appendix~\ref{app:type-II}, shows that the most
important vertex diagrams are those represented in Fig.~\ref{fig:fig11}. Diagram
(II.R1) describes the correlated state of three atoms during the rf conversion.
Before the conversion, the atom in state $|2\rangle$ is entangled with an atom
in state $|1\rangle$. This entanglement is preserved once the atom $|2\rangle$
is excited by the rf radiation to state $|3\rangle$. If the interaction $V_{12}$
is attractive, this process enhances the effect of the final-state interaction
$V_{13}$ by increasing the probability that the excited atom has an atom in
state $|1\rangle$ nearby. If $V_{12}$ is repulsive, this process keeps the
excited atom away from atoms in state $|1\rangle$, reducing the effect of
$V_{13}$. The effect of the final-state interaction $V_{23}$, on the other hand,
is limited by the Pauli principle: Just after the conversion, the atom
$|3\rangle$ sits in the correlation hole of the former atom $|2\rangle$ and is
kept away from other atoms in state $|2\rangle$. The exclusion principle indeed
forbids any contribution like (II.R1), where the atom $|1\rangle$ would be
replaced by an atom $|2\rangle$. The converted atom can nevertheless interact
with atoms in state $|2\rangle$, either directly (self-energy corrections of the
$3p$ line) or via the exchange process represented by the diagram (II.R2). In
this process, the converted atom interacts with an atom in state $|2\rangle$
that is present above the Fermi energy, such that the interaction does not
produce a new hole. The atom $|2\rangle$ eventually recombines with the hole
left by the conversion, while the atom $|3\rangle$ is converted back to an atom
$|2\rangle$ above the Fermi sea.

In the experimental setup of Ref.~\onlinecite{Frohlich-2012}, the scattering
length $a_{12}$ measuring the interaction $V_{12}$ between states $|1\rangle$
($|F=9/2,m_F=-9/2\rangle$) and $|2\rangle$ ($|F=9/2,m_F=-7/2\rangle$) is close
to a Feshbach resonance and was tuned on the attractive side from $0$ to $-300$,
in units of the Bohr radius $a_{\text{B}}$. The interactions $V_{13}$ and
$V_{23}$ between states $|1\rangle$ and $|3\rangle$ ($|F=9/2,m_F=-5/2\rangle$)
and $|2\rangle$ and $|3\rangle$ are both nonresonant and repulsive and
correspond to scattering lengths $a_{13}=+250a_{\text{B}}$ and
$a_{23}=+130a_{\text{B}}$. In this configuration, we expect that the effect of
$V_{13}$ is enhanced by the attractive $V_{12}$ in the correction (II.R1), while
$V_{23}$ only contributes through the exchange process (II.R2). We speculate
that most of the extra broadening observed in the measurements, with respect to
the theory including $V_{12}$, but neglecting final-state interactions
\cite{Frohlich-2012}, is the result of these processes. A self-energy broadening
due to the direct interaction between $|2\rangle$ and $|3\rangle$ in the initial
state (accounted for in the type-I diagram of Fig.~\ref{fig:fig03}) is unlikely
because such contributions require at least two holes in the final state and are
suppressed by a factor $e^{-h\nu_0/k_{\text{B}}T}$ (see
Appendix~\ref{app:type-II}). If the number of atoms excited in the final state
is not too small, the self-energy in the final state associated with both
$V_{13}$ and $V_{23}$ may also induce, in addition to energy shifts at lowest
order, some broadening of type I, which enters the resolution function, as shown
in Sec.~\ref{sec:final-state1}. Explicit evaluations of the vertex corrections
in Fig.~\ref{fig:fig11} and of other self-energy effects are left for future
works. It will be interesting to see whether and how these final-state effects
change the line shape of the rf signal.

\section{Conclusion}
\label{sec:conclusion}

We have presented a theoretical description of the rf spectroscopy of cold-atom
systems, based on the second-order response theory at finite temperature. The
difference between the usual golden-rule approach and this new description is
that the latter focuses on the number $N_f$ of atoms transferred to the final
state, while the former focuses on the transition rate $\dot{N}_f$. The
second-order response approach accounts for the finite energy resolution implied
by the envelope and the finite duration of the rf pulse and allows one to
classify the various contributions using Feynman diagrams. The issue of
inhomogeneity represents a challenge for the interpretation of rf experiments
performed on interacting Fermi systems. Due to the density dependence of the
self-energy, the rf line shape varies across the cloud. We have studied this
effect at leading order in the density within the LDA and found that the static
local Hartree shifts induce an apparent dispersion of the rf signal, similar to
the dispersion expected in a homogeneous interacting Fermi gas from dynamical
effects of higher order in the density. For three-dimensional gases with a
Gaussian density profile, we have derived a simple expression for the
momentum-integrated rf line shape, which takes into account the finite
resolution and the inhomogeneous Hartree shifts.

Final-state effects are another challenge for rf experiments. We have considered
the simplest of them, resulting either from a lifetime or from the interplay of
inhomogeneity and Hartree shifts in the final state. More subtle final-state
effects, such as those resulting from the spatial correlations between atoms,
are described by vertex corrections. We have proposed a scheme to classify these
terms and identified those which dominate at low temperature and low density. A
numerical evaluation of the corresponding diagrams is needed to tell whether
these effects change significantly the line shape of the rf signal.

\acknowledgments

We acknowledge useful discussions with D. S. Jin. This work was supported by the
Swiss National Science Foundation under Division II, the Alexander-von-Humboldt
Stiftung, and the European Research Council (Grant No. 616082).

\appendix

\section{Analytic continuation of second-order response functions}
\label{app:Matsubara-double-time}

In this appendix, we show that the second-order retarded susceptibility, defined
in terms of the double commutator in Eq.~(\ref{eq:nk2}), corresponds by
analytical continuation to the imaginary-time correlator
(\ref{eq:C2definition}). We switch to a slightly lighter notation, set
$\hbar=1$, and compute the second-order change of the expectation value of an
observable $A$, in the presence of a perturbation $H'=BF(t)$, where $B$ is an
observable and $F(t)$ is a classical field. The second-order correction is
	\begin{multline}
		\langle A(t)\rangle^{(2)}=(-i)^2\int_{-\infty}^tdt_1
		\int_{-\infty}^{t_1}dt_2\\ \times\langle[[A(t),H'(t_1)],H'(t_2)]\rangle_H.
	\end{multline}
The ensemble average is taken over the eigenstates of the time-independent
Hamiltonian $H$, such that invariance by translation in time applies:
$\langle[[A(t),H'(t_1)],H'(t_2)]\rangle_H =
\langle[[A(t-t_1),H'(0)],H'(t_2-t_1)]\rangle_H$. Using this, and the expression
of $H'$, we can write
	\begin{equation}\label{eq:app1}
		\langle A(t)\rangle^{(2)}=\int_{-\infty}^{\infty}dt_1dt_2\,
		\chi_{AB}^{(2)}(t-t_1,t-t_2)F(t_1)F(t_2),
	\end{equation}
with the second-order susceptibility defined as
	\begin{multline}\label{eq:app11}
		\chi_{AB}^{(2)}(t,t')=(-i)^2\theta(t)\theta(t'-t)\\
		\times \langle[[A(t),B(0)],B(t-t')]\rangle_H.
	\end{multline}
Introducing the Fourier transform of the various quantities in the integrand of
Eq.~(\ref{eq:app1}) leads to the analogous of the second line in
Eq.~(\ref{eq:nk2}):
	\begin{multline}
		\langle A(t)\rangle^{(2)}=\int_{-\infty}^{\infty}\frac{d\omega}{2\pi}
		\frac{d\omega'}{2\pi}\,e^{-i(\omega+\omega')t}\\
		\times \chi_{AB}^{(2)}(\omega,\omega')F(\omega)F(\omega').
	\end{multline}
Out task is to show that
	\begin{equation}\label{eq:app2}
		\chi_{AB}^{(2)}(\omega,\omega')=\frac{1}{2}
		\mathscr{C}_{AB}^{(2)}(i\Omega\to\omega+i0^+,i\Omega'\to\omega'+i0^+),
	\end{equation}
where $\mathscr{C}_{AB}^{(2)}(i\Omega,i\Omega')$ is the Fourier transform of the
imaginary-time correlator
	\begin{equation}
		\mathscr{C}_{AB}^{(2)}(\tau,\tau')=\langle T_{\tau}A(\tau)B(0)B(\tau-\tau')\rangle_H.
	\end{equation}
For this purpose, we show that the spectral representations of the functions
$\chi_{AB}^{(2)}(\omega,\omega')$ and
$\frac{1}{2}\mathscr{C}_{AB}^{(2)}(i\Omega,i\Omega')$ are identical.

Let us start with $\mathscr{C}_{AB}^{(2)}$. Splitting the imaginary-time
integrals to take into account the time ordering, we have
	\begin{align}\label{eq:app3}
		\nonumber
		&\mathscr{C}_{AB}^{(2)}(i\Omega,i\Omega')\\
		\nonumber
		&\qquad=\int_0^{\beta}d\tau d\tau'\,
		e^{i\Omega\tau}e^{i\Omega'\tau'}\mathscr{C}_{AB}^{(2)}(\tau,\tau')\\
		\nonumber
		&\qquad=\int_0^{\beta}d\tau\,e^{i\Omega\tau}
		\left[
		\int_0^{\tau}d\tau'\,e^{i\Omega'\tau'}\langle A(\tau)B(\tau-\tau')B(0)\rangle_H
		\right.\\ &\quad\qquad \left.
		+\int_{\tau}^{\beta}d\tau'\,e^{i\Omega'\tau'}\langle A(\tau)B(0)B(\tau-\tau')\rangle_H
		\right].
	\end{align}
To perform the time integrations, we introduce a complete set of eigenstates of
$H$, $H|a\rangle=E_a|a\rangle$, we use the expression of the thermal average,
$\langle(\cdots)\rangle=Z^{-1}\text{Tr}\,e^{-\beta H}(\cdots)$, we insert two
times the identity $\openone=\sum_a|a\rangle\langle a|$, and we use the
expression of the imaginary-time operators, e.g., $A(\tau)=e^{\tau H}Ae^{-\tau
H}$. The averages in the square brackets of (\ref{eq:app3}) become
	\begin{align*}
		&\langle A(\tau)B(\tau-\tau')B(0)\rangle_H\\
		&\qquad=\frac{1}{Z}\sum_{abc}\langle a|A|b\rangle
		\langle b|B|c\rangle\langle c|B|a\rangle
		e^{E_a(\tau-\beta)}e^{-E_b\tau'}e^{E_c(\tau'-\tau)}\\
		&\langle A(\tau)B(0)B(\tau-\tau')\rangle_H\\
		&\qquad=\frac{1}{Z}\sum_{abc}\langle a|A|b\rangle
		\langle b|B|c\rangle\langle c|B|a\rangle
		e^{E_a(\tau'-\beta)}e^{-E_b\tau}e^{E_c(\tau-\tau')}.
	\end{align*}
The $\tau$ and $\tau'$ integrations in (\ref{eq:app3}) are now elementary and
yield, after making use of the property $e^{i\Omega\beta}=e^{i\Omega'\beta}=1$,
	\begin{multline}\label{eq:app4}
		\mathscr{C}_{AB}^{(2)}(i\Omega,i\Omega')=\frac{1}{Z}\sum_{abc}
		\langle a|A|b\rangle\langle b|B|c\rangle\langle c|B|a\rangle\\
		\times\frac{1}{i\Omega+i\Omega'+E_a-E_b}\left(
		\frac{e^{-\beta E_a}-e^{-\beta E_c}}{i\Omega+E_a-E_c}
		+\frac{e^{-\beta E_b}-e^{-\beta E_c}}{i\Omega-E_b+E_c}\right.\\ \left.
		+\frac{e^{-\beta E_a}-e^{-\beta E_c}}{i\Omega'+E_a-E_c}
		+\frac{e^{-\beta E_b}-e^{-\beta E_c}}{i\Omega'-E_b+E_c}
		\right).
	\end{multline}
A similar calculation leads to the spectral representation of the real-time
susceptibility. We start from
	\begin{multline}\label{eq:app5}
		\chi_{AB}^{(2)}(\omega,\omega')=-\int_{-\infty}^{\infty}dtdt'\,
		e^{i\omega t}e^{i\omega't'}\theta(t)\theta(t'-t)\\
		\times\langle[[A(t),B(0)],B(t-t')]\rangle_H.
	\end{multline}
The four terms of the double commutator are expressed as
	\begin{align*}
		&\langle A(t)B(0)B(t-t')\rangle_H\\
		&\qquad=\frac{1}{Z}\sum_{abc}\langle a|A|b\rangle
		\langle b|B|c\rangle\langle c|B|a\rangle
		e^{-\beta E_a}e^{iE_at'}e^{-iE_bt}e^{iE_c(t-t')}\\
		&\langle B(t-t')A(t)B(0)\rangle_H\\
		&\qquad=\frac{1}{Z}\sum_{abc}\langle a|A|b\rangle
		\langle b|B|c\rangle\langle c|B|a\rangle
		e^{-\beta E_c}e^{iE_at'}e^{-iE_bt}e^{iE_c(t-t')}\\
		&\langle B(0)A(t)B(t-t')\rangle_H\\
		&\qquad=\frac{1}{Z}\sum_{abc}\langle a|A|b\rangle
		\langle b|B|c\rangle\langle c|B|a\rangle
		e^{-\beta E_c}e^{iE_at}e^{-iE_bt'}e^{iE_c(t'-t)}\\
		&\langle B(t-t')B(0)A(t)\rangle_H\\
		&\qquad=\frac{1}{Z}\sum_{abc}\langle a|A|b\rangle
		\langle b|B|c\rangle\langle c|B|a\rangle
		e^{-\beta E_b}e^{iE_at}e^{-iE_bt'}e^{iE_c(t'-t)}.
	\end{align*}
We perform the time integrations in (\ref{eq:app5}) with the help of the identity
	\[
		\int_{-\infty}^{\infty}dt\,e^{i\omega t}\theta(t)=\frac{i}{\omega+i0^+},
	\]
and obtain, using the notations $\omega^+=\omega+i0^+$ and
${\omega'}^+=\omega'+i0^+$,
	\begin{multline*}
		\chi_{AB}^{(2)}(\omega,\omega')=\frac{1}{Z}\sum_{abc}
		\langle a|A|b\rangle\langle b|B|c\rangle\langle c|B|a\rangle\\
		\times\frac{1}{\omega^++{\omega'}^++E_a-E_b}\left(
		\frac{e^{-\beta E_a}-e^{-\beta E_c}}{{\omega'}^++E_a-E_c}
		+\frac{e^{-\beta E_b}-e^{-\beta E_c}}{{\omega'}^+-E_b+E_c}\right).
	\end{multline*}
By exchanging the dummy variables $t_1$ and $t_2$ in the expression
(\ref{eq:app1}), we see that the susceptibility (\ref{eq:app11}) can also be
defined with the arguments $t$ and $t'$ exchanged. We could therefore use an
alternate definition of the susceptibility, which shows explicitly the symmetry
under the exchange of the time arguments, e.g.,
$\frac{1}{2}[\chi^{(2)}_{AB}(t,t')+\chi^{(2)}_{AB}(t',t)]$ instead of
Eq.~(\ref{eq:app11}). Exchanging the time arguments in Eq.~(\ref{eq:app11}) is
equivalent to exchanging the two frequencies $\omega$ and $\omega'$ in
Eq.~(\ref{eq:app5}). After performing this symmetrization, we obtain the
alternate definition of the susceptibility:
	\begin{multline}\label{eq:app6}
		\chi_{AB}^{(2)}(\omega,\omega')=\frac{1}{2}\frac{1}{Z}\sum_{abc}
		\langle a|A|b\rangle\langle b|B|c\rangle\langle c|B|a\rangle\\
		\times\frac{1}{\omega^++{\omega'}^++E_a-E_b}\left(
		\frac{e^{-\beta E_a}-e^{-\beta E_c}}{\omega^++E_a-E_c}
		+\frac{e^{-\beta E_b}-e^{-\beta E_c}}{\omega^+-E_b+E_c}\right. \\ \left.
		+\frac{e^{-\beta E_a}-e^{-\beta E_c}}{{\omega'}^++E_a-E_c}
		+\frac{e^{-\beta E_b}-e^{-\beta E_c}}{{\omega'}^+-E_b+E_c}
		\right).
	\end{multline}
Comparison of Eqs.~(\ref{eq:app4}) and (\ref{eq:app6}) proves
(\ref{eq:app2}).

\section{Momentum density and density distribution}
\label{app:nkD}

By inverting the analog of Eq.~(\ref{eq:self-consistent-Hartree}) for
$n_2(\vec{r})$, one obtains an expression for
$\mu-(1/2)m\omega_r^2r^2-(g/N_0)n_1(\vec{r})$ as a function of $n_2(\vec{r})$.
Inserting this expression into Eq.~(\ref{eq:nktgLDA}) gives
	\begin{equation*}
		\langle n_{\vec{k}}\rangle_{\text{LDA}}=\int d^dr\,
		\frac{\mathscr{R}^{\text{(I$'$)}}
		\big(h\tilde{\nu}-(g/N_0)n_1(\vec{r})\big)}
		{1-e^{\beta\varepsilon_{\vec{k}}}\Big/\text{Li}_{d/2}^{-1}\left[
		-\left(\frac{2\pi\hbar^2}{mk_{\text{B}}T}\right)^{d/2}n_2(\vec{r})\right]},
	\end{equation*}
where $\text{Li}_{n}^{-1}$ is the inverse of the polylogarithm function. If
$n_1(\vec{r})=n_2(\vec{r})\equiv n(\vec{r})$, the $\vec{r}$-dependence of the
integrand stems from $n(\vec{r})$, and the spatial integration can be converted
into a density integration, by introducing the density distribution
$D(\mathfrak{n})=\int d^dr\,\delta\big(\mathfrak{n}-n(\vec{r})\big)$:
	\begin{equation}\label{eq:nkgLDA_D}
		\langle n_{\vec{k}}\rangle_{\text{LDA}}=\int_{-\infty}^{\infty}d\mathfrak{n}\,
		\frac{D(\mathfrak{n})\mathscr{R}^{\text{(I$'$)}}
		\big(h\tilde{\nu}-(g/N_0)\mathfrak{n}\big)}
		{1-e^{\beta\varepsilon_{\vec{k}}}\Big/\text{Li}_{d/2}^{-1}\left[
		-\left(\frac{2\pi\hbar^2}{mk_{\text{B}}T}\right)^{d/2}\mathfrak{n}\right]}.
	\end{equation}
For an ideal resolution,
$\mathscr{R}^{\text{(I$'$)}}(\varepsilon)\propto\delta(\varepsilon)$, we have
simply
	\begin{equation*}
		\langle n_{\vec{k}}\rangle_{\text{LDA}}\propto\frac{D(h\tilde{\nu}N_0/g)}
		{1-e^{\beta\varepsilon_{\vec{k}}}\Big/\text{Li}_{d/2}^{-1}\left[
		-\left(\frac{2\pi\hbar^2}{mk_{\text{B}}T}\right)^{d/2}
		\frac{h\tilde{\nu}N_0}{g}\right]}.
	\end{equation*}
This expression can be made more explicit in dimension $d=2$. On the one hand,
$\text{Li}_1^{-1}(x)=1-e^{-x}$, and on the other hand, the density distribution
can be evaluated explicitly. We have
	\begin{equation*}
		D(\mathfrak{n})=\begin{cases}\displaystyle
		\frac{2\pi r_0}{|n'(r_0)|} & 0\leqslant \mathfrak{n}\leqslant n(0)\vspace{2mm}\\
		0 & \text{otherwise},\end{cases}
	\end{equation*}
where $n'(r)$ is the derivative of the radial density $n(r)$, $n(0)$ is the
density at the trap center, and $n(r_0)=\mathfrak{n}$. Differentiating
Eq.~(\ref{eq:self-consistent-Hartree}) with respect to $r$, one finds
	\begin{equation*}
		\frac{2\pi r}{|n'(r)|}=\frac{2\pi}{m\omega_r^2N_0}\left[1+g+e^{-
		\frac{\mu-\frac{1}{2}m\omega_r^2r^2-(g/N_0)n(r)}
		{k_{\text{B}}T}}\right].
	\end{equation*}
The exponential in the square brackets can be expressed as a function of $n(r)$
only, by inverting Eq.~(\ref{eq:self-consistent-Hartree}) as above. For $r=r_0$,
on thus gets
	\begin{equation*}
		\frac{2\pi r_0}{|n'(r_0)|}=\frac{2\pi}{m\omega_r^2N_0}\left[1+g+
		b\left(\frac{\mathfrak{n}}{N_0}\right)\right],
	\end{equation*}
where $b(\varepsilon)=1/(e^{\varepsilon/k_{\text{B}}T}-1)$. The resulting
expression for the momentum distribution in two dimensions, and for an ideal
resolution, is given in Eq.~(\ref{eq:nkgLDA0}). Interestingly, the functional
dependence of the density distribution on $\mathfrak{n}$, and consequently the
dependence of the momentum distribution (\ref{eq:nkgLDA0}) on $\tilde{\nu}$,
does not involve the total particle number $N_1$; only the cutoff depends on
$N_1$ via $n(0)$.

\section{Classification of vertex corrections}
\label{app:type-II}

\begin{figure}[tb]
\includegraphics[width=0.8\columnwidth]{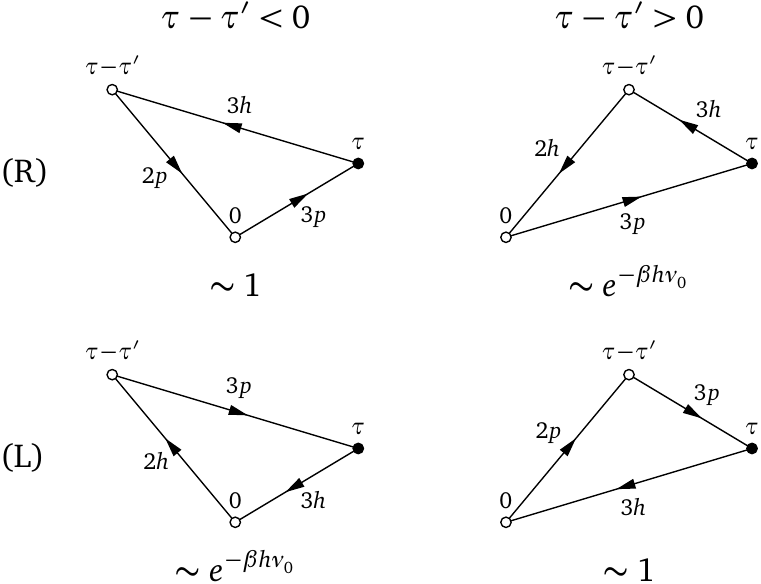}
\caption{\label{fig:fig12}
Contributions to the right-handed (R) and left-handed (L) diagrams of zeroth
order in the interaction. The vertices are ordered horizontally by increasing
imaginary time from left to right. $2p$ and $3p$ indicate particle lines in
states $|2\rangle$ and $|3\rangle$, respectively, while $2h$ and $3h$ indicate
hole lines.
}
\end{figure}

The upper line in the diagram of type R in Fig.~\ref{fig:fig02} corresponds to a
hole in the final state $|3\rangle$, as implied by the ordering of the times,
e.g., $\tau-\tau'<\tau$. The lower line corresponds to a particle in the final
state. Conversely, in the diagram of type L, the lower line corresponds to a
hole ($0<\tau$) and the upper line to a particle. In both cases, the vertical
line describes either a particle or a hole in the initial state $|2\rangle$,
depending upon the ordering of the times $\tau-\tau'$ and $0$. This is
illustrated in Fig.~\ref{fig:fig12} in the case of type-I diagrams. Each hole in
the state $|3\rangle$ entails an occupation factor $f(\varepsilon_3)\sim
e^{-\beta h\nu_0}$, which is negligible if the thermal population of the final
state is negligible. One such factor is canceled if---and only if---the time
$\tau'-\tau$ can reach the value $\beta$. (This applies to R diagrams; the same
statement with $\tau'-\tau$ replaced with $\tau-\tau'$ applies to L diagrams.)
The reason is as follows. The Green's function for a free hole propagating
between times $\tau_1$ and $\tau_2$ is
$f(\varepsilon_3)e^{-\varepsilon_3(\tau_2-\tau_1)}$, whereas for a free particle
it is $-f(-\varepsilon_3)e^{-\varepsilon_3(\tau_2-\tau_1)}$. All time
dependencies from the various particle and hole lines in state $|3\rangle$
cancel, except at the two conversion vertices ($\circ$), leaving only the
dependence $e^{-\varepsilon_3(\tau-\tau')}$. Upon performing the time
integrations as specified by Eq.~(\ref{eq:C2Fourier}), a factor $e^{\beta
h\nu_0}$ is generated if the time $\tau'-\tau$ ($\tau-\tau'$ for L diagrams) is
allowed to reach the value $\beta$. This explains the behaviors indicated in
Fig.~\ref{fig:fig12}. In all cases, there is one hole in the final state ($3h$
line)---hence a factor $e^{-\beta h\nu_0}$---that is canceled for right-handed
diagrams if $\tau-\tau'<0$ and for left-handed ones if $\tau-\tau'>0$. The two
types of contributions were denoted (Ia) and (Ib) in
Sec.~\ref{sec:leading_contribution}.

\begin{figure}[b]
\includegraphics[width=\columnwidth]{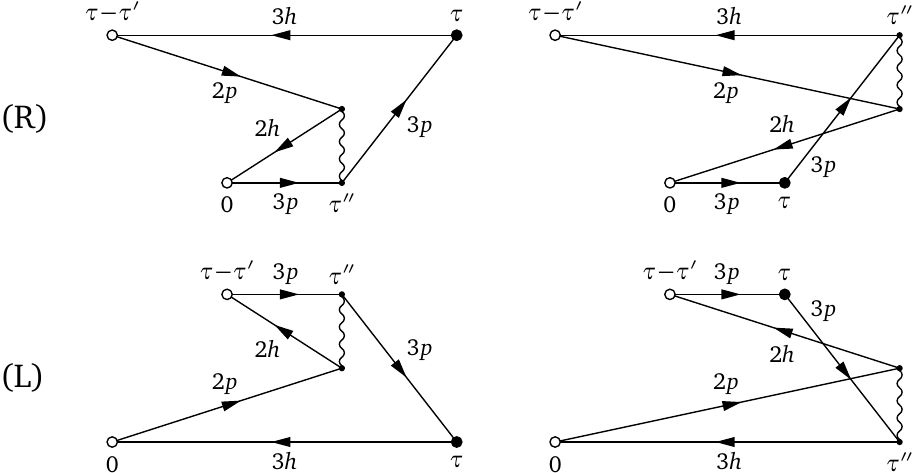}
\caption{\label{fig:fig13}
First-order right-handed (R) and left-handed (L) contributions to the
$\circ$-vertex corrections, which survive in the limit $k_{\text{B}}T\ll
h\nu_0$. The imaginary times $\tau-\tau'$, $0$, $\tau''$, and $\tau$ are ordered
horizontally as in Fig.~\ref{fig:fig12}. Any modification in the ordering of
times produces at least one factor $e^{-\beta h\nu_0}$.
}
\end{figure}

Since the cancellation of the final-state hole occupation factor can only work
once, we conclude that any diagram with more than one hole in the state
$|3\rangle$ carries at least one factor $e^{-\beta h\nu_0}$ and is exponentially
small if $k_{\text{B}}T\ll h\nu_0$. In particular, all corrections of the
density vertex ($\bullet$) imply a connection between the lines $3p$ and $3h$
that cuts the $3h$ line and thus contains at least two holes in the final state.
The first-order corrections of the conversion vertices ($\circ$) which survive
in the limit $k_{\text{B}}T\ll h\nu_0$ are displayed in Fig.~\ref{fig:fig13}.

At higher orders in the interaction, we classify the vertex corrections like in
the low-density expansion of the self-energy \cite{Galitskii-1958}. A
self-energy diagram containing $p$ hole lines, for instance a particle-hole
ladder at order $p+1$, is proportional to $e^{p\beta\mu}$. Since $\mu\to-\infty$
as the density $n\to0$ at any finite temperature, the contributions with one
single hole dominate in this limit. These contributions are given by the
particle-particle ladder series. Similarly, the vertex corrections with one
single hole in either of the initial states $|1\rangle$ or $|2\rangle$ are
expected to dominate at low density. Figure~\ref{fig:fig11} shows the two
contributions which we consider as the most important vertex corrections at low
density. Both contain a single hole in the final state, and a single hole in one
of the initial states. Any further decoration of these diagrams with interaction
lines introduces new hole lines. The two right-handed first-order terms of
Fig.~\ref{fig:fig13} may be obtained from the diagram (II.R2) by removing one of
the interaction boxes and evaluating at first order.

\end{document}